\documentclass[a4paper,11pt]{article}%
\usepackage{geometry}
\usepackage{amsmath}
\usepackage{amsfonts}
\usepackage{amssymb}
\usepackage{graphicx}
\usepackage{indentfirst}
\usepackage{hyperref}

\usepackage{latexsym}

\usepackage[small,bf]{caption}
\usepackage{slashed}
\usepackage{braket,euscript}
\usepackage{fancyvrb}
\usepackage{amsmath,amssymb,url}
\providecommand{\U}[1]{\protect\rule{.1in}{.1in}}
\newcommand{\bea}{\begin{eqnarray}}	
\newcommand{\eea}{\end{eqnarray}}
\newcommand{\be}{\begin{equation}}	
\newcommand{\ee}{\end{equation}}
\newcommand{\beq}{\begin{equation}}	
\newcommand{\eeq}{\end{equation}}

\newcommand{\Z}{{\mathbb Z}}
\newcommand{\C}{{\mathbb C}}

\def\R{\relax\ifmmode {\mathbb R}  \else${\mathbb R}$\fi}
\def\C{\relax\ifmmode {\mathbb C}  \else${\mathbb C}$\fi}
\def\Z{\relax\ifmmode {\mathbb Z}  \else${\mathbb Z}$\fi}
\def\N{\relax\ifmmode {\mathbb N}  \else${\mathbb N}$\fi}
\def\I{\relax\ifmmode {\mathbb I}  \else${\mathbb I}$\fi}

\begin{document}

\date{}
\title{\textbf{Study of the all orders multiplicative renormalizability of a local matter confining Gribov-Zwanziger action in the MAG}}
\author{\textbf{D.~Fiorentini%$^1$
}\thanks{diego\_fiorentini@id.uff.br}  \\[2mm] %\,\,,
%\textbf{V.~J.~Vasquez~Otoya$^2$}\thanks{victor.vasquez@ifsudestemg.edu.br}\\[2mm]
{\small \textnormal{ %$^1$
\it Instituto de F\'isica, Universidade Federal Fluminense, UFF - }}
 \\ \small \textnormal{ \it Campus da Praia Vermelha, Niter\'oi, 24210-340, RJ, Brazil.} \\\\
 %{\small \textnormal{$^2$\it IFSEMG $-$ Instituto Federal de Educac\~ao, Ci\^encia e Tecnologia,}}\\\small\textnormal{ \it Rua Bernardo Mascarenhas 1283, 36080-001, Juiz de Fora, MG, Brasil}
 \normalsize}

\maketitle

\begin{abstract}
 
We address the issue of the all order multiplicative renormalizability of $SU(2)$ Gribov-Zwanziger theories quantized in the maximal Abelian gauge in presence of confined matter fields. The non-linear character of the maximal Abelian gauge requires the introduction of quartic interaction terms in the Faddeev-Popov ghosts as well as in the localizing Zwanziger fields, extended a well known feature of this gauge.  We show that, when scalar matter fields are introduced,  a second quartic interaction term in the scalar fields, Faddeev-Popov ghosts and Zwanziger-like fields naturally arises. A BRST invariant action accounting for those quartic interaction terms was identified and it was proven to be multiplicative renormalizable to all orders by means of the algebraic renormalization procedure.  
%%%%%%%%%%%%%%%%%%%%%%%%%%%%%%%%%%%%%%%%%%%%%%%%%%
\end{abstract}

\section{Introduction}

One of most important problems of modern physics is how to explain the confinement of color charged particles  
inside hadrons, \textit{i.e.}, the experimental fact that quarks and gluons have not been  detected in isolation, but exclusively as constituents of hadrons. Nowadays, the maximal Abelian gauge \cite{'tHooft:1981ht,Kronfeld:1987vd,Kronfeld:1987ri} is widely employed in  order to investigate such a phenomena by means of isolation of physical relevant parameters in the infrared sector, namely this gauge turns out to be suitable for the study of the dual superconductivity mechanism for color confinement \cite{Mandelstam:1974pi,Nambu:1974zg,tHooft:1975krp}, according to which $SU(N)$ Yang-Mills theories  region should be described as an effective $U(1)^{N-1}$ Abelian theory in low-energy \cite{Ezawa:1982bf,Suzuki:1989gp,Suzuki:1992gz,Hioki:1991ai,Sakumichi:2014xpa} in the presence of monopoles. The condensation of these magnetic charges leads to a dual Meissner effect, where the QCD vacuum behave as a 
superconductor of chromo-magnetic current, compressing field lines of chromo-electric charge in a 
flux tube (string), resulting in quark confinement by generating a linear interquark potential, as Abrikosov string is formed in a Cooper pairs medium. In particular, to avoid unnecessary complications, when restricted to $SU(2)$ YM theory,  the Abelian configuration is identified with the diagonal components $A_{\mu }^{3}$ of the gauge field 
corresponding to the diagonal generator of the Cartan subgroup of $SU(2)$. The
remaining off-diagonal components $A_{\mu }^{\alpha}$, $\alpha=1,2$, corresponding to the off-diagonal generators of $SU(2)$, are expected to acquire a mass through a dynamical mechanism, thus decoupling at low energies. This phenomenon is known as Abelian dominance and it is object of intensive investigation, both from analytic and from numerical lattice simulations \cite{Sasaki:1998ww,Sasaki:1998th,Suzuki:1995va,Suzuki:1989gp,Shiba:1994ab,Stack:1994wm,Hioki:1991ai,Miyamura:1995xn}. A considerable amount of evidence to the dynamical mass generation for the off-diagonal components of the gauge field from the analytic side can be found in \cite{Schaden:1999ew,Kondo:2001nq,Dudal:2004rx}, while \cite{Amemiya:1998jz,Bornyakov:2003ee,Gongyo:2013sha} are devoted to numerical studies.
 
Besides being a renormalizable gauge  \cite{Min:1985bx,Fazio:2001rm,Gracey:2005vu}, the maximal Abelian gauge enjoys the important property  of exhibiting a lattice formulation \cite{Amemiya:1998jz,Bornyakov:2003ee,Gongyo:2013sha,Mendes:2006kc,Mihari:2007zz}, a property which allow us to compare analytic and numerical results. In particular, this property has made possible the study, from the numerical lattice point of view, of the behaviour of the two-point gluon correlation function in the non-pertutbative infrared region, thus providing us evidence of the Abelian dominance as well as of the confining character of the propagator of the  Abelian gluon component \cite{Amemiya:1998jz,Bornyakov:2003ee,Gongyo:2013sha,Mendes:2006kc,Mihari:2007zz}. This issue has also been addressed through analytical methods by taking into account the existence of the Gribov copies  \cite{Gribov:1977wm} which, as in any covariant and renormalizable gauge,  affect the maximal Abelian gauge \cite{Bruckmann:2000xd,Guimaraes:2011sf,Capri:2013vka}. Here, by proceeding in a similar way to the Landau gauge \cite{Sobreiro:2005ec,Vandersickel:2012tz}, some properties of the so-called Gribov region have been derived and the restriction of the integration domain in the functional integral to the Gribov horizon was formulated, see for instance refs.\cite{Capri:2006cz,Capri:2008ak,Capri:2008vk,Capri:2010an} for the details of the Gribov problem in the maximal Abelian gauge.  Remarkably, the agreement between the lattice numerical results and the analytic calculations based on the restriction to the Gribov region looks quite good \cite{Mendes:2006kc,Capri:2008ak}, confirming the expectation that the study of the Gribov problem is of great relevance for gluon confinement.

Nevertheless, so far, the analytical study of the infrared aspects in the maximal Abelian gauge has been done only in the gluon sector, without including matter fields, {\it i.e.} spinor and scalar fields. This work focus at continuing the analytic study of the non-perturbative behaviour of the  matter fields in the maximal Abelian gauge started in \cite{Capri:2015pxa,Capri:2017abz}, along the lines recently outlined in  \cite{Dudal:2008sp,Dudal:2011gd,Capri:2014bsa} for the Landau gauge case, where it has been possible to recover the behaviour of the propagators for scalar and spinor fields observed in lattice simulations in \cite{Maas:2011yx,Maas:2010nc,Furui:2006ks,Parappilly:2005ei} from an analytic point of view \cite{Capri:2014bsa}. This study is relevant in the MAG context because it would allow us to extend the Abelian dominance hypotesis to matter sector and thus make predictions about  the propagator of scalars and quark fields which might be compared to lattice numerical simulations \cite{Schrock:2015pna}.

The first step in this endeavour was performed in \cite{Capri:2015pxa}, where the all orders multiplicative renormalizability of the $SU(2)$ Yang-Mills theory fixed in the maximal Abelian gauge in presence of matter fields was established, a topic which, till then, has not yet been addressed.  The goal of the present paper is to extend this proof to the case in which the Gribov problem  \cite{Gribov:1977wm} is taking into account and a Gribov-like confinement mechanism for matter fields is implemented. Although the renormalizability of the maximal Abelian gauge in presence of the matter fields is an expected feature, we had noted that it is not a straightforward matter, requiring in fact a  nontrivial analysis. This is due to the non-linear character of the maximal Abelian gauge which gives rise to a rather complex Faddeev-Popov operator. It was already pointed out that the structure of this operator requires the introduction of a quartic interaction between ghosts \cite{Min:1985bx,Fazio:2001rm,Gracey:2005vu}. Only at the very end of the whole renormalization process the gauge parameter entering the quartic interaction can be set to zero \cite{Min:1985bx,Fazio:2001rm,Gracey:2005vu}, thus recovering the genuine  maximal Abelian gauge condition. In \cite{Capri:2015pxa}, we had seen that this feature generalizes to the case of scalar matter fields, {\it i.e.} a quartic interaction term between scalar fields and Faddeev-Popov ghosts naturally arises due to the non-linearity of the gauge condition. As a consequence, a second gauge parameter associated to this new term has to be introduced. As in the case of the quartic ghost term, this second gauge parameter can be set to zero only at the very end of the renormalization process, since it is not a physical (observable) quantity.

%%%%%%%%%%%%%%%%%%%%%%%%%%%%%%%%%%%%%%%%%%%%%%%%%%%%
%%%%%%%%%%%%%%%%%%%%%%%%%%%%%%%%%%%%%%%%%%%%%%%%%%%%
%%%%%%%%%%%%%%%%%%%%%%%%%%%%%%%%%%%%%%%%%%%%%%%%%%%%
Concerning to the Gribov problem in the maximal Abelian gauge, although the situation cannot  be compared to that of the Landau gauge \cite{Gribov:1977wm,Zwanziger:1988jt,Zwanziger:1989mf,Zwanziger:1992qr,Dell'Antonio:1991xt}, a few results are already available, see \cite{Capri:2005tj,Capri:2006cz,Capri:2007hw,Capri:2008ak,Capri:2008vk,Capri:2008ak,Capri:2010an,Capri:2011ki}, where an analogous of Zwanziger horizon function  as well as of the  Gribov-Zwanziger action and of its refined version have been constructed. A study of the maximal Abelian gauge within the context of the Schwinger-Dyson equations can be found in \cite{Huber:2009wh}. Gribov analysis starts noting that Faddeev-Popov quantization program is incomplete, because equivalent gauge fields configurations survive to gauge-fixing procedure as consequence of the presence of zero modes of the Faddeev-Popov operator \cite{Bruckmann:2000xd,Guimaraes:2011sf,Capri:2013vka}, which is given by \eqref{offop} for the $SU(2)$ symmetry group\footnote{Notations and definitions are given in the Sec. \ref{MatterConfMod}.}. By restricting the domain of integration in functional Feynman integral to the so-called Gribov region, where Faddeev-Popov operator is strictly positive, a large number of copies could be eliminated, as proved in \cite{Capri:2005tj}. In complete analogy to Landau case, can be proved that this restriction is equivalent to adding to the original Faddeev-Popov action the called horizon term \cite{Capri:2010an} (see the next section for details), which is proportional to the massive parameter $\gamma$, called Gribov parameter\footnote{This is a dynamical quantity determined by gap equation  as function of coupling constant and scale invariant  \cite{Capri:2005tj}, being suppressed in UV sector. Thus, horizon functions affects the non-perturbative infrared behavior of gluedynamics only.}. Although horizon term is non-local, it can be cast in local form by following the Zwanziger method, namely by introducing a quartet of the so-called Zwanziger fields, resulting in a renormalizable Gribov-Zwanziger action \cite{Capri:2006cz}. As in Landau gauge \cite{Dudal:2008sp,Dudal:2011gd}, these auxiliary fields develope a non-trivial dynamics as consequence of infrared non-vanishing value of Gribov parameter \cite{Capri:2008ak}. In this scenario, off-diagonal gluons should acquire a sufficiently large dynamical mass which decouples them at low energy due to the dynamical mass generation coming from the condesation of the dimension two gluon operator $A_{\mu}^\alpha A_{\mu}^\alpha$ \cite{Kondo:2001nq,Dudal:2004rx}, \textit{i.e.}, $\langle A_{\mu}^\alpha A_{\mu}^\alpha\rangle\sim m^2$, and its propagator turns out to be of the Yukawa type. Another two condensates arise for ghost fields and auxiliary localizing fields . Dynamical mass associated with one of this new condensates,  $\langle\bar{\varphi}\varphi-\bar{\omega}\omega-\bar{c}c\rangle\sim\mu^2$, introduce modifications of the Gribov-Stingl type in the diagonal gluon propagator, which attain 
non-vanishing value at zero momentum, $k^2=0$ and it does not exhibit the K\"all\'en-Lehmann spectral 
representation, so that it cannot be associated with the propagation of physical particles. This fact means 
that diagonal gluons are not physical excitations of the theory. Then, refinement GZ framework by 
taking into account the condensation of two dimensional operators with MAG fixing condition for 
pure Yang-Mills theories  confirm the confining character of diagonal gluon propagators, as 
predicted by lattice simulations \cite{Amemiya:1998jz,Bornyakov:2003ee,Mendes:2006kc,Mendes:2008ux,Gongyo:2012jb,Gongyo:2013sha}.  In this way, the Abelian dominance conjecture can be considerer as the analytical confining criterion emerging from fundamental Yang-Mills theory  by taking into account the existence of Gribov copies when maximal Abelian gauge fixing condition is used in the quantization procedure of the theory. 

In this work, we use an extension of this statement for the case in which matter fields are included, following the proposal in \cite{Capri:2014bsa} for Landau gauge, according to which Faddeev-Popov operator coupling in universal way to any color charged field. Thus, for all generic matter field $F^i$ in a given representation of $SU(N)$ (with Latin indexes) specified by the generators $(T^a)^{ij}$, a non-local term,  similar to the horizon function, should be added to full action (\textit{vide} also \cite{Palhares:2016wqn}), namely: 
\begin{equation}
H_{matter}(G)=Gg^2\int d^4xd^4yF^i(x)(T^a)^{ij}(\mathcal{M}^{-1})^{ab}(x,y)(T^b)^{jk}F^k(y)
\label{Hmatter1}
\end{equation}
where $G$ plays a role akin to that of the Gribov parameter for  matter sector.   In MAG case, for adjoint representation, since Faddeev-Popov operator contains off-diagonal color indexes only, just the diagonal matter fields can be coupling to it, as happens for gauge fields case. In this sense, propagators should show Abelian dominance. In fundamental representation, there is no qualitative distinction with the propagator in Landau case reported in \cite{Capri:2014bsa}. Moreover, in the same way that  Landau case, localizing Zwanziger fields for horizon matter term \eqref{Hmatter1} develop a non-trivial dynamics associated to new Gribov-like parameter for horizon matter term which modify the  infrared behaviour of the propagators.

%%%%%%%%%%%%%%%%%%%%%%%%%%%%%%%%%%%%%%%%%%%%%%%%%%
%%%%%%%%%%%%%%%%%%%%%%%%%%%%%%%%%%%%%%%%%%%%%%%%%%
%%%%%%%%%%%%%%%%%%%%%%%%%%%%%%%%%%%%%%%%%%%%%%%%%%
It is necessary to emphasize that recently in \cite{Capri:2015ixa,Capri:2016aqq,Capri:2015pxa} was established BRST invariant formulation of the Gribov–Zwanziger theory which is local, albeit nonpolynomial; in particular, for the maximal Abelian gauge, see \cite{Capri:2015pxa}. This construction has allowed a geometrical resolution of the Gribov problem in the class of the linear covariant gauges, however, for the specific cases of Landau gauge and MAG, the invariant Gribov-Zwanziger formulation correspond to rewrite the Nakanishi-Lautrup sector in terms of a infinite power series of the divergence of the gauge field, which is nothing but a trivial change of variables in the path integral because the Jacobian of the transformation is unitary \cite{Capri:2015ixa,Capri:2016aqq,Capri:2015pxa}. From this point of view,  both the BRST-exact and no-exact formulations are equivalent via an adequate  reparametrization, \textit{ergo} they share the same formal properties.. In particular, in \cite{Capri:2017abz},  a BRST-exact horizon-like matter term was implemented following the aforementioned statement.

The  aim of the present work is to prove the important fact that a new horizon term as \eqref{Hmatter1} in matter sector does not spoil the renormalizability of the theory. In \cite{Capri:2008ak} an all order multiplicative renormizability proof for SU(2) pure Yang-Mills theories is performed when the Gribov problem is taking into account. The inclusion of Zwanziger localizing  fields for the horizon term are controlled by the existence of a new class of symmetries in these fields, allowing us to define a large set of Ward identities which guaranteed the algebraic renormalization proof. Moreover, in a previous work \cite{Capri:2015pxa} it was proved that Yang-Mills theories remain renormalizable when self-interacting scalar matter minimally coupled to gauge field is present. Inclusion of horizon term should not modify UV-sector since it just introduce low-energy effects only.  In this paper we show that a new symmetry arise too for the localizing fields of horizon term for matter,  guaranteeing the existence of a large set of Ward identities in a similar manner to the identities involving the Zwanziger fields for the gauge sector. This fact will be taken as sufficient evidence that the renormalizability is not jeopardized by the inclusion of a confining horizon term, while the BRST-exact and no-exact equivalence via the Nakanishi-Lautrup reparametrization  will be taken as sufficient evidence of the renormalization of the BRST-exact formulation given in \cite{Capri:2017abz}.

The present work is organized as follows. In Sect. 2 we briefly discuss the maximal Abelian gauge and the corresponding gauge fixing as well as the Gribov-Zwanziger formulation when infrared effects are taking into account and when we include the scalar matter sector. The Sect.3  is devoted to the extension of the physical action in order to restore the BRST symmetry and we elaborate on the quartic interactions required to renormalizes the theory. Sect.4 is devoted to establish the set of Ward identities needed for the all orders proof of the renormalizability and we present the algebraic characterization of the most general invariant local counterterm, and we prove the renormalizability of model to all orders by means of the algebraic renormalization. Sect. 5 collects our conclusion, whereas the Appendix A show as to generalize the previous renormalization arguments to the fermion matter case.  

%%%%%%%%%%%%%%%%%%%%%%%%%%%%%%%%%%%%%%%%%
%%%%%%%%%%%%%%%%%%%%%%%%%%%%%%%%%%%%%%%%%
\section{Matter confinement model}
\label{MatterConfMod}
%%%%%%%%%%%%%%%%%%%%%%%%%%%%%%%%%%%%%%%%%
%%%%%%%%%%%%%%%%%%%%%%%%%%%%%%%%%%%%%%%%%

\subsection{Maximal Abelian gauge condition and Gribov problem}

To avoid unnecessary complications and with no loss of generality, for the rest of this paper,  we restrict ourselves to the case of the gauge group $SU(2)$. In order to fix the notation for the Yang-Mills action in the maximal Abelian gauge (MAG), we start by considering a $SU(2)$-Lie algebra valued gauge field $\mathcal{A}_{\mu}=A^a_\mu T^a$, where the algebra generators $T^a\,\;(a=1,..,3)$ is given by
\begin{equation}
\left[ T^a,T^b\right] =i\varepsilon ^{abc}T^c  \label{la}
\end{equation}
and they are chosen to be anti-Hermitean and to obey the orthonormality condition $\,
\mathrm{Tr}\left( T^aT^b\right) =\delta ^{ab}$. 
Following  \cite{'tHooft:1981ht,Kronfeld:1987vd,Kronfeld:1987ri}, the gauge field can be decomposed into diagonal and off-diagonal components, namely
\begin{equation}
\mathcal{A}_{\mu}=A_\mu ^\alpha T^\alpha +A_\mu T^{\,3}  \label{cd}
\end{equation}
where $\alpha=1,2$ and $T^3\equiv T$ is the diagonal generator of the Cartan subgroup of $SU(2)$. Thus, the  commutation relations \eqref{la} adopts the form
\begin{eqnarray}
\left[T^\alpha,T^\beta\right]&=&i\epsilon^{\alpha\beta 3}T^3\equiv i\epsilon^{\alpha\beta}T\,,\nonumber\\
\left[T^\alpha,T\right]&=&-i\epsilon^{\alpha\beta}T^\beta\,,\quad \left[T,T\right]=0
\label{mag2}
\end{eqnarray}
where $\epsilon^{\alpha\beta}= \epsilon^{\alpha\beta 3}$. Analogously, we can express the Yang-Mills action as
\begin{equation}
S_{\mathrm{YM}}=\frac{1}{4}\int d^dx\left(F^{\alpha}_{\mu\nu}F^{\alpha}_{\mu\nu}+F_{\mu\nu}F_{\mu\nu}\right)
\label{mag5}
\end{equation}
by using the following explicit field strength decomposition
\begin{eqnarray}
F^{\alpha}_{\mu\nu} &=& \mathcal{D}^{\alpha\beta}_\mu A^{\beta}_\nu-\mathcal{D}^{\alpha\beta}_\nu A^{\beta}_\mu\nonumber\\
F_{\mu\nu} &=& \partial_\mu A_\nu - \partial_\nu A_\mu + g\epsilon^{\alpha\beta}A^{\alpha}_\mu A^{\beta}_\nu
\label{mag3}
\end{eqnarray}
with $\mathcal{D}^{\alpha\beta}_\mu$ being the covariant derivative defined with respect to the Abelian component\footnote{From here, we use alway the notation without colour super-index, $A_\mu=A_\mu^3$, for Abelian components.}, namely
\begin{equation}
\mathcal{D}^{\alpha\beta}_\mu=\delta^{\alpha\beta}\partial_\mu-g\epsilon^{\alpha\beta}A_{\mu}
\label{mag4}
\end{equation}
and which is left invariant under the following infinitesimal gauge transformations,
\begin{subequations}
\begin{align}
\delta A^{\alpha}_\mu &= -\mathcal{D}^{\alpha\beta}_\mu \xi^\beta - g\epsilon^{\alpha\beta}A^{\beta}_{\mu}\xi 
\\
\delta A_\mu &= -\partial_\mu \xi - g\epsilon^{\alpha\beta}A^{\alpha}_\mu \xi^\beta
\end{align}
\label{mag6}
\end{subequations} 
The maximal Abelian gauge condition amounts to impose that the off-diagonal
components $A_{\mu }^{\alpha}$ of the gauge field obey the following nonlinear condition
\begin{equation}
\mathcal{D}_{\mu }^{\alpha\beta}A_{\mu }^{\beta}=0  \label{offgauge}
\end{equation}
which follows by requiring that the auxiliary functional
\begin{equation}
\mathcal{R}[A]=\int {d^{4}x}A_{\mu }^{\alpha}A_{\mu }^{\alpha}  \label{fmag}
\end{equation}
be stationary with respect to the gauge transformations (\ref{mag6}).
Moreover, as it is apparent from the presence of the covariant derivative $\mathcal{D}_{\mu }^{\alpha\beta}$, equation \eqref{offgauge} allows for a residual local $U(1)$
invariance corresponding to the diagonal subgroup of $SU(2)$. This
additional invariance has to be fixed by means of a further gauge condition
on the diagonal component $A_{\mu }$, which is usually chosen to be of the
Landau type, namely
\begin{equation}
\partial _{\mu }A_{\mu }=0\;.  \label{dgauge}
\end{equation}
The Faddeev-Popov operator, $\mathcal{M}^{\alpha\beta}$, corresponding to the gauge condition (\ref{offgauge}) is easily derived by taking the second variation of the auxiliary functional $\mathcal{R}[A]$, being given by
\begin{equation}
\mathcal{M}^{\alpha\beta}=-\mathcal{D}_{\mu }^{\alpha\gamma}\mathcal{D}_{\mu }^{\gamma\beta}-g^{2}\varepsilon
^{\alpha\gamma}\varepsilon ^{\beta\omega}A_{\mu }^{\gamma}A_{\mu }^{\omega}  \label{offop}
\end{equation}
It enjoys the property of being Hermitian and, as pointed out in \cite{Bruckmann:2000xd}, is the difference of two positive semidefinite operators given by $-\mathcal{D}_{\mu }^{\alpha\omega}\mathcal{D}_{\mu }^{\omega\beta}$ and $g^{2}\varepsilon^{\alpha\omega}\varepsilon ^{\beta\rho}A_{\mu }^{\omega}A_{\mu }^{\rho}$,  respectively. It is worth to point out that the operator $\mathcal{M}^{\alpha\beta}$ is non-linear in the gauge fields, a feature which has  nontrivial consequences in the renormalization process both in the case of the gauge and matter sector. The gauge fixed Yang-Mills action in the MAG is written as
\begin{equation}
S^{\mathrm{FP}}_{\mathrm{MAG}} = S_{\mathrm{YM}}+\int d^4x\left\{b^\alpha \mathcal{D}^{\alpha\beta}_\mu A^\beta_\mu - \bar{c}^{\alpha}\mathcal{M}^{\alpha\beta}c^\beta + g\epsilon^{\alpha\beta}\bar{c}^{\alpha}(\mathcal{D}^{\alpha\delta}_{\mu}A^{\delta}_{\mu})c+b\partial_\mu A_\mu +\bar{c}\partial_\mu (\partial_\mu c + g\epsilon^{\alpha\beta}A^\alpha_\mu c^\beta)\right\}
\label{mag9}
\end{equation}
where $({\bar c}^{\alpha}, {\bar c}, c^{\alpha}, c)$ are the Faddeev-Popov ghosts and $(b^{\alpha},b)$ are the Lagrange multipliers implementing the gauge conditions \eqref{offgauge} and \eqref{dgauge}. %Further, we introduce the  $s$-exact gauge fixing term
%%%%%%%%%%%%%%%%%%%%%%%%%%%%%%%%%%%%%%%%%%%%%%%%%%%%%%%%%%%%
%%%%%%%%%%%%%%%%%%%%%%%%%%%%%%%%%%%%%%%%%%%%%%%%%%%%%%%%%%%%
%%%%%%%%%%%%%%%%%%%%%%%%%%%%%%%%%%%%%%%%%%%%%%%%%%%%%%%%%%%%
%%%%%%%%%%%%%%%%%%%%%%%%%%%%%%%%%%%%%%%%%%%%%%%%%%%%%%%%%%%%

As any other covariant gauge, also the maximal Abelian gauge is plagued by the existence of Gribov copies, \textit{vide} refs.\cite{Bruckmann:2000xd,Guimaraes:2011sf,Capri:2013vka} for explicit examples of zero modes of the Faddeev-Popov operator \eqref{offop}.  An analogous of the Gribov region of the Landau gauge can be introduced in the MAG by restricting the integration in the functional  integral to the region where the Faddeev-Popov operator $\mathcal{M}^{\alpha \beta}$ is strictly positive, {\it i.e.} $\mathcal{M}^{\alpha \beta}>0$, then a large number of copies could be eliminated, as proven in \cite{Capri:2005tj,Capri:2008vk}. Furthermore, in complete analogy with the case of the Landau gauge, this restriction can be implemented in a path integral formulation by adding to the original Faddeev-Popov action \eqref{mag9}  a non-local horizon term which, in the case of the maximal Abelian gauge, turns out to be given by the expression \cite{Capri:2005tj,Capri:2006cz,Capri:2008ak,Capri:2008vk}
\begin{equation}
H_{MAG}(A)= g^2\int d^4x\;d^4y\;A_{\mu}(x)\varepsilon^{\alpha \beta}\left(\mathcal{M}^{-1}\right)^{\alpha \delta}(x,y)\varepsilon^{\delta \beta}A_{\mu}(y)
\label{H_MAG}
\end{equation}
Therefore, we have  the analogous of the Gribov-Zwanziger action in the maximal Abelian gauge as given by
\begin{equation}
S^{\text{GZ}}_{\text{MAG}} = S_{\text{MAG}}^{\text{FP}} + \gamma^4 H_{\text{MAG}}(A)
\label{gzmag}
\end{equation}
where $\gamma^2$ stands for the Gribov parameter of the maximal Abelian gauge. By proceeding as in the case of the Landau gauge, the expression \eqref{gzmag} can be cast in local form by introducing a pair of auxiliary boson fields, $(\bar{\varphi}^{\alpha \beta}_{\mu}, {\varphi}^{\alpha \beta}_{\mu})$, and a pair of auxiliary fermion fields, $(\bar{\omega}^{\alpha \beta}_{\mu}, {\omega}^{\alpha \beta}_{\mu})$, namely\footnote{Note the presence of the term
%%%%%%%%%%%%%%%%%%%%%%%%%%%%%%
\begin{equation}
\mathcal{F}^{\alpha\beta}=2g\varepsilon^{\alpha\gamma}\left(\partial_{\mu}c+g\varepsilon^{\delta\omega}A_{\mu}^\delta c^\omega\right)D_{\mu}^{\gamma\beta}
+g\varepsilon^{\alpha\beta}\partial_\mu\left(\partial_{\mu}c+g\varepsilon^{\gamma\delta}A_{\mu}^\gamma c^\delta\right)
-g^2(\varepsilon^{\alpha\gamma}\varepsilon^{\beta\delta}-\varepsilon^{\alpha\delta}\varepsilon^{\beta\gamma})A_{\mu}^\delta \left( D_\mu^{\gamma\rho} c^\rho + g\varepsilon^{\gamma\rho}A_\mu^\rho c\right)
\end{equation}
in \eqref{locgzmag}, which come from the trivial shift $\omega^{\alpha \beta}_\mu \to \omega^{\alpha \beta}_\mu - \mathcal{M}^{-1,\alpha\delta}\mathcal{F}^{\delta\gamma}\varphi^{\gamma\beta}_\mu$, whose corresponding Jacobian is field-independent, and it is necessary to write the local  $\gamma-$independent horizon term in a BRST-exact form, due to the transformation of the hidden $A_\mu$ field in the covariant derivative including in the Faddeev-Popov operator. 
%%%%%%%%%%%%%%%%%%%%%%%%%%%%%%%%%%%%%%%%%5
}
\begin{equation} 
S^{\text{GZ}}_{\text{MAG}} = S_{\text{MAG}}^{\text{FP}} + \int d^4x\left\{\bar{\varphi}^{\alpha \beta}_{\mu}\mathcal{M}^{\alpha \delta}\varphi^{\delta \beta}_{\mu}
-\bar{\omega}^{\alpha \beta}_\mu \mathcal{M}^{\alpha \delta}\omega^{\delta \beta }_\mu
+\bar{\omega}^{\alpha \beta}_\mu \mathcal{F}^{\alpha \delta}\varphi^{\delta \beta }_\mu
 + g\gamma^2 \varepsilon^{\alpha \beta}\left(\varphi -\bar{\varphi} \right)^{\alpha \beta}_\mu A_{\mu}  \right\} \label{locgzmag}
\end{equation} 
As shown in \cite{Capri:2005tj,Capri:2006cz,Capri:2008ak,Capri:2008vk}, the action $S^{\text{GZ}}_{\text{MAG}}$ enables us to implement the restriction in the functional integral to the Gribov region $\Omega_{\text{MAG}}$ of the maximal Abelian gauge, defined as 
\begin{equation}
\Omega_{\text{MAG}} = \left\{ \; A^\alpha_\mu\,, A_{\mu}\;\;\left| \;  \;  \partial_\mu A_\mu =0,  \; D^{\alpha \beta}_\mu A^{\beta}_\mu=0,  \;    {\mathcal M}^{\alpha \beta}(A) =  -\left( D_{\mu }^{\alpha \delta}D_{\mu }^{\delta \beta}+g^{2}\varepsilon
^{\alpha \sigma}\varepsilon ^{\beta \delta }A_{\mu }^{\sigma}A_{\mu }^{\delta} \right)> 0  \;\right. \right\}   
\label{om}
\end{equation}
Although the understanding of the Gribov issue in the maximal Abelian gauge cannot yet be compared to that reached in the Landau gauge, a few properties of the region $\Omega_{\text{MAG}}$ have been already obtained. In particular, in  \cite{Capri:2008vk}, it has been established that $\Omega_{\text{MAG}}$ is unbounded along the diagonal directions in field space. This feature seems to be consistent with the aforementioned Abelian dominance hypothesis, according to which the diagonal  configurations, corresponding to the Abelian Cartan subgroup, should be the dominant configurations in the infrared. Moreover, in  \cite{Greensite:2004ke}, it has been shown that when an Abelian configuration is gauge-transformed to the Landau gauge, it is mapped into a point of the boundary of the Gribov region $\Omega$ of the Landau gauge, eq.\eqref{om}, {\it i.e.} into a point of the Gribov horizon\footnote{See Sect.V of  \cite{Greensite:2004ke}.}. These features give further support to the restriction of the domain of integration to the  region $\Omega_{\text{MAG}}$. 
%%%%%%%%%%%%%%%%%%%%%%%%%%%%%%%%%%%%%%%%%%%%%%%%%%%%
%%%%%%%%%%%%%%%%%%%%%%%%%%%%%%%%%%%%%%%%%%%%%%%%%%%%

We can now address the issue of the  existence of a nilpotent non-perturbative BRST symmetry for the action \eqref{locgzmag}. In \cite{Capri:2015pfa}, a non-local expression  has been used  to built-up a  gauge-invariant gluon field configuration, in terms of which an invariant horizon function was introduced in the MAG by generalizing the construction proposed in \cite{Capri:2015ixa} for the Landau and linear covariant gauges. Thus, an infrared modified BRST operator can be inferred for this new formulation: we may say that the BRST transformation \textit{feels} the restriction of the path integral to the Gribov region too. As it was turn out, the construction of the transverse gauge-invariant field $A_\mu^h$ follows from  the minimization  of the Hilbert norm along the gauge orbit of a given configuration $A_{\mu }^a$, which correspond to the minimal of the functional
\begin{equation}
f_{A}[U]\equiv \mathrm{Tr}\int d^{4}x\,A_{\mu }^{U}A_{\mu}^{U}
\label{fa}
\end{equation}
where we shall require that both $A_{\mu }^{a}$ and the local gauge transformations, $U\in SU(N)$, are square-integrable, as showed in \cite{Zwanziger:1990tn, Lavelle:1995ty,Capri:2005dy,Capri:2015ixa}. Making  use of this configurations, we can rewrite expression  \eqref{H_MAG} as 
\begin{eqnarray} 
H_{\text{MAG}}(A) &= & H_{\text{MAG}}(A^h) - {\mathcal F}(A) \partial A -  {\mathcal F}^{\alpha}(A) \partial A^{\alpha} \nonumber \\
& = &  H_{\text{MAG}}(A^h) - \left[ - \partial_\mu {\mathcal F}(A) + g \varepsilon^{\alpha \beta}  {\mathcal F}^{\alpha}(A) A^\beta_\mu \right] A_\mu - {\mathcal F}^{\alpha}(A) D^{\alpha \beta}_\mu A^\beta_\mu 
\label{Hh}
\end{eqnarray} 
where we are using the short-hand notation  ${\mathcal F}(A) (\partial A)= \int d^4x d^4y {\mathcal F}(x,y) (\partial A)_y$ and  ${\mathcal F}^{\alpha}(A) (\partial A^{\alpha})= \int d^4x d^4y {\mathcal F}^{\alpha}(x,y) (\partial A^\alpha)_y$, and ${\mathcal F}(A)$ stands for an infinite non-local power series of $A_\mu^a$\footnote{Mathematical features are not relevant at the present discussion, since the change of variables implemented so far has a illustrative character; the reference \cite{Capri:2015pfa} is remitted for technical details about the $A^h$ construction.}. The \textit{residual} terms in ${\mathcal F}(A)$ can be reabsorbed by a harmless shift of the fields $(b^\alpha,b)$; namely, by introducing  the redefined Lagrange multipliers $b^h, b^{h,\alpha}$ as following
\begin{eqnarray} 
b^h & = & b -\gamma^4 {\mathcal F}(A) + \gamma^4 \int_{-\infty}^{x} dy_\mu \left(  g \varepsilon^{\alpha \beta}  {\mathcal F}^{\alpha}(A) A^\beta_\mu \right)_y \nonumber \\
b^{h,\alpha} & = & b^{\alpha} - \gamma^4 {\mathcal F}^{\alpha}(A)    \label{redb}
\end{eqnarray}
we can rewrite the action \eqref{gzmag} as 
\begin{equation}
S^{\text{GZ}}_{\text{MAG}} = S^{\text{FP}}_{\text{MAG}}(b^h, b^{h,\alpha})  + \gamma^4 H_{MAG}(A^h) 
 \label{gzmag1}
\end{equation}
which enables us to write down an exact nilpotent non-perturbative BRST symmetry \cite{Capri:2015pfa}. Notice that equations \eqref{redb}  correspond to a linear change of variables in the path integral in the $b$-sector of the theory,  thus corresponding to a trivial Jacobian.

%%%%%%%%%%%%%%%%%%%%%%%%%%%%%%%%%%%%%%%%%%%%%%%%%%%%%%%
The aim of the last analysis is that of establishing a equivalence between the Gribov-Zwanziger actions \eqref{gzmag} and  \eqref{gzmag1} via the reparametrizations \eqref{Hh} and \eqref{redb}. As  previously mentioned, it is means that both the actions share the same physical properties, and, in particular, the renormalizability of the one implies that of the other. The action \eqref{gzmag} suffers from a BRST soft symmetry breaking  in the conventional context of the GZ framework, but can be embedded into a larger action so that, following the general lines of the procedure proposed in \cite{Zwanziger:1992qr}, the BRST symmetry can be restored by means of a suitable set of external sources. This process has the advantage of being much simpler than to implement the algebraic renormalization for a nonpolynomial model like \eqref{gzmag1}, where the renormalization factors can be nonlinear, \textit{i.e.} they can be power series, as in the linear covariant gauge case\footnote{Note that at LCG case we are forced to use the nonpolynomial form in order to implement the Gribov-Zwanziger framework due to the loss of Hermiticity of the Faddeev-Popov.} \cite{Capri:2017bfd}. Of course, we avoid unnecessary complications by using the polynomial form \eqref{gzmag} in the MAG case with no loss of generality.

%%%%%%%%%%%%%%%%%%%%%%%%%%%%%%%%%%%%%%%%%%%%%%%%%%

In the next Subsection, we proceed to add scalar matter following the conjecture in \cite{Capri:2014bsa} of universal coupling of the  Faddeev-Popov operator to any (BRST-invariant) coloured field in order to universalize the confining character of the Gribov-Zwanziger construction.
%%%%%%%%%%%%%%%%%%%%%%%%%%%%%%%%%%%%%%%%%%%%%%%%%%%%%%%

\subsection{Confinement action for scalar matter in the adjoint representation}
\label{SubSecConfScalarMatter}

To begin, we consider the case in which $SU(2)-$valued scalar matter  is present, $\mathcal{\phi}= \phi^a T^a$,  then we add a Klein-Gordon action with self-interaction in the adjoint  representation minimally coupled to gauge field with the full covariant derivative, \textit{i.e.}, the  coupling term in the covariant derivative include diagonal and off-diagonal components of gauge  field. For consistency, we perform the Cartan decomposition of matter field too, {\it i.e.}  we decompose the scalar field into off-diagonal and diagonal components, 
\begin{equation}
\phi^aT^a=\phi^\alpha T^\alpha+ \phi T^{\,3}  \label{sf}
\end{equation}
The corresponding matter action is given by
\begin{eqnarray}
S_{\text{scalar}}&=&\int d^4x \left(\frac{1}{2}(D_{\mu}^{ab}\phi ^b)^2+\frac{m^2_{\phi}}{2}\phi^a\phi^a+\frac{\lambda}{4!}(\phi ^a\phi ^a)^2\right)\nonumber\\
&=&\int d^4x \left\{(\partial_{\mu}\phi^\alpha)(\partial_{\mu}\phi^\alpha)+(\partial_{\mu}\phi)(\partial_{\mu}\phi)
-2g^2\varepsilon^{\alpha\beta}\left[(\partial_{\mu}\phi)\phi^\alpha A_{\mu}^\beta
-(\partial_{\mu}\phi^\alpha)\phi A_{\mu}^\beta+(\partial_{\mu}\phi^\alpha)\phi^\beta A_{\mu}\right]
\right.\nonumber\\
&&+g^2\left[A_{\mu}^\alpha A_{\mu}^\alpha\left(\phi^\beta\phi^\beta+\phi\phi\right)
+A_{\mu}A_{\mu}\phi^\alpha\phi^\alpha -A_{\mu}^\alpha A_{\mu}^\beta\phi^\alpha \phi^\beta
-2A_{\mu}^\alpha A_{\mu}\phi^\alpha \phi\right]
\nonumber\\
&&\left.+\frac{m^2_{\phi_{off}}}{2}\phi^\alpha\phi^\alpha +\frac{m^2_{\phi_{diag}}}{2}\phi\phi
+\frac{\lambda}{4!}\left[\left(\phi^\alpha\phi^\alpha\right)^2+2\phi^2\phi^\alpha\phi^\alpha+\phi^4\right]\right\}
\label{smatter}
\end{eqnarray}
The classical action $(S_{\text{YM}}+S_{\text{scalar}})$ is left invariant
by the gauge transformations
\begin{eqnarray}
\delta A_{\mu }^{\alpha} &=&-D_{\mu }^{\alpha\beta}{\omega }^{\beta}-g\varepsilon ^{\alpha\beta}A_{\mu
}^{\beta}\omega  \nonumber \\
\delta A_{\mu } &=&-\partial _{\mu }{\omega }-g\varepsilon ^{\alpha\beta}A_{\mu
}^{\alpha}\omega ^{\beta} \label{gauge}
\end{eqnarray}
and 
\begin{equation}
\delta \phi^{\alpha}=g\varepsilon^{\alpha\beta}\phi\,\omega ^{\beta}-g\varepsilon ^{\alpha\beta}\phi ^\beta \omega \,,\qquad \delta \phi=-g\varepsilon^{\alpha\beta}\phi^a\omega ^\beta  \label{sct} 
\end{equation} 

%%%%%%%%%%%%%%%%%%%%%%%%%%%%%%%%%%%%%%%%%%%%%%%%%%%%%%%%%%%%%%%%%
%%%%%%%%%%%%%%%%%%%%%%%%%%%%%%%%%%%%%%%%%%%%%%%%%%%%%%%%%%%%%%%%%
%%%%%%%%%%%%%%%%%%%%%%%%%%%%%%%%%%%%%%%%%%%%%%%%%%%%%%%%%%%%%%%%%
%%%%%%%%%%%%%%%%%%%%%%%%%%%%%%%%%%%%%%%%%%%%%%%%%%%%%%%%%%%%%%%%%

In  complete analogy with the pure gluon sector and following the general prescription proposed in \cite{Capri:2014bsa}, a non-local non-perturbative term must be included for matter. By using \eqref{Hmatter1}, the confining character for the matter sector is implemented by adding a horizon-like term in the matter sector where the inverse of Faddeev-Popov operator is coupled only to Abelian component of scalar field in the adjoint representation
\begin{equation}
S_{\text{matter}}=S_{\text{scalar}}+ \sigma^4H_{\text{matter}}(\phi)
\end{equation} 
where 
\begin{equation}
H_{\text{matter}}(\phi)=g^2\int d^4x\,d^4y\,\epsilon ^{\alpha\beta}\phi(x)(\mathcal{M}^{-1})^{\alpha\gamma}(x,y)\epsilon ^{\gamma\beta}\phi(y)
\label{H_MAG_scalar}
\end{equation}

In this case, the parameter $\sigma$ would be the Gribov-like parameter for matter sector, in analogy to gauge sector case. The horizon-type function for matter \eqref{H_MAG_scalar} can be cast in local form by introducing one pairs of bosonic fields, $(\bar{\eta},\eta)^{\alpha\beta}_{\mu}$, and other of anti-commutating fields, $(\bar{\theta},\theta)^{\alpha\beta}_{\mu}$: 
\begin{eqnarray}
H_{\text{matter}}^{(\text{local})}&=&S_{\eta\theta}+S_{\sigma}\nonumber\\
&=&\int d^4x\left\{\bar{\eta}^{\alpha\beta}\mathcal{M}^{\alpha\gamma}\eta^{\gamma\beta}
-\bar{\theta}^{\alpha\beta}\mathcal{M}^{\alpha\gamma}\theta^{\gamma\beta}+\bar{\theta}^{\alpha\gamma}\mathcal{F}^{\alpha\beta}\eta^{\beta\gamma}
+\sigma^2 g\varepsilon^{\alpha\beta}\left(\eta-\bar{\eta}\right)^{\alpha\beta}\phi
\right\}
\label{S0}
\end{eqnarray}
Finally, the complete physical action is given by
\begin{eqnarray}
S_{\text{phys}}&=&S_{\text{YM}}+S_{\text{MAG}}+S_{\varphi\omega}+S_{\gamma}+S_{\text{matter}}+S_{\eta\theta}+S_{\sigma}
\nonumber\\
&=&\lim_{\alpha,\xi\rightarrow0}\int d^4x \left\{\frac{1}{4}\left(F_{\mu\nu}^{\alpha}\right)^2
+\frac{\alpha}{2}\left(b^\alpha\right)^2 +\frac{\xi}{2}b^2+b^\alpha\mathcal{D}_{\mu}^{\alpha\beta}A_{\mu}^\beta  +b\partial_{\mu}A_{\mu}
-\bar{c}^\alpha\mathcal{M}^{\alpha\beta}c^\beta 
+g\varepsilon^{\alpha\beta}\bar{c}^\alpha\left(\mathcal{D}_{\mu}^{\beta\omega}A_{\mu}^\omega\right)c
\right.\nonumber\\
&&+\bar{c}\partial_{\mu}\left(\partial_{\mu}c+g\varepsilon^{\alpha\beta}A_{\mu}^\alpha c^\beta\right)
+\bar\varphi^{\alpha\omega}_\mu\mathcal{M}^{\alpha\beta}\varphi^{\beta\omega}_\mu
-\bar\omega^{\alpha\omega}_\mu\mathcal{M}^{\alpha\beta}\omega^{\beta\omega}_\mu
+g\gamma^2\varepsilon^{\alpha\beta}\left(\varphi-\bar{\varphi}\right)^{\alpha\beta}_{\mu} A_{\mu}
\nonumber\\
&&\left. 
+\left(\mathcal{D}_{\mu}^{ab}\phi^b\right)^2
+\frac{m^2}{2}\phi^a\phi^a
+\frac{\lambda}{4!}\left(\phi^a\phi^a\right)^2
+\bar{\eta}^{\alpha\omega}\mathcal{M}^{\alpha\beta}\eta^{\beta\omega}
-\bar{\theta}^{\alpha\omega}\mathcal{M}^{\alpha\beta}\theta^{\beta\omega}
+g\sigma^2\varepsilon^{\alpha\beta}\left(\eta-\bar{\eta}\right)^{\alpha\beta}\phi\right\}
\label{Sphys1}                  
\end{eqnarray}
The BRST variations of the Faddeev-Popov fields can be logically extended for all the Zwanziger(-type) localizing fields to remains nilpotent ($s^2=0$)
\begin{eqnarray}
sA^{\alpha}_{\mu}=-(D^{\alpha\beta}_{\mu}c^{\beta}+g\varepsilon^{\alpha\beta}A_{\mu}^\beta c)\,,\qquad sA_{\mu}=-(\partial_{\mu}c+g\varepsilon^{\alpha\beta}A_{\mu}^\alpha c^\beta)\nonumber\\ 
sc^{\alpha}=g\varepsilon^{\alpha\beta}c^{\beta}c\,,\qquad sc=\frac{g}{2}\varepsilon^{\alpha\beta}c^{\alpha}c^\beta \,,\qquad  s\bar{c}^{\alpha}=b^{\alpha}\,,\qquad s\bar{c}=b\,,\qquad
sb^{\alpha}=sb=0 &\nonumber\\
s\phi^{\alpha}=g\varepsilon^{\alpha\beta}(\phi\,c^{\beta}-\phi ^\beta c)\,,\qquad s\phi=-g\varepsilon^{\alpha\beta}\phi^\alpha c^\beta\,&\nonumber\\ 
s\bar\omega^{\alpha}_{I}=\bar\varphi^{\alpha}_{I}\,,\qquad s\bar\varphi^{\alpha}_{I}=0\,,\qquad
s\varphi^{\alpha}_{I}=\omega^{\alpha}_{I}\,,\qquad s\omega^{\alpha}_{I}=0\,&\nonumber\\
s\bar\theta^{\alpha}_i=\bar\eta^{\alpha}_i\,,\qquad s\bar\eta^{\alpha}_i=0\,,\qquad
s\eta^{\alpha}_i=\theta^{\alpha}_i\,,\qquad s\theta^{\alpha}_i=0&
\label{BRSTscalar}
\end{eqnarray} 

%We need to do two remarks at this point. First, the doublet structure in the BRST transformations above  for the Zwanziger(-type) localizating fields are required to guaranteed their cohomological triviality \textit{w.r.t.} the BRST operator \cite{Zwanziger:1992qr}; the (soft) breaking problem related to this choice will be study in the following section in the context of the renormalization procedure. Second

\section{Extended BRST-exact form of physical action for scalar matter}  

Now, let's start the study of the renormalizability of the of full physical action \eqref{Sphys1}. In order to achieve this aim, the first step is note that \eqref{Sphys1} exhibits a soft breaking of the BRST symmetry due to the presence of the Gribov (and Gribov-like) parameter(s), namely
\begin{equation}
sS_{\text{phys}}=-g\varepsilon^{\alpha\beta}\int d^4x
\left\{\gamma^2\left[\varphi^{\alpha\beta}_{\mu}(\partial_{\mu}c+g\varepsilon^{\gamma\omega}A_{\mu}^\gamma c^\omega)-\omega^{\alpha\beta}_{\mu}A_{\mu}\right]
+\sigma^2\left[\eta^{\alpha\beta}(g\varepsilon^{\gamma\omega}\phi^\gamma c^\omega)-\theta^{\alpha\beta}\phi\right]
\right\}
\label{SoftBreak}
\end{equation}
where $s$ stands for the nilpotent BRST transformations \eqref{BRSTscalar}. In order to restore the BRST symmetry, we embed \eqref{Sphys1} into a larger BRST-exact action, following the method introduced by D. Zwanziger in \cite{Zwanziger:1992qr}. The detailed construction of this general action follows the same steps  as described in \cite{Capri:2006cz,Capri:2008ak,Capri:2014bsa,Capri:2015pxa,Capri:2014fsa} and we need to keep in mind that the physical action is re-obtained from the extended action, which will be denote as $\Sigma$, when the set of external sources and parameters attain their physical values, {\it i.e.}
\begin{equation}
\left.\Sigma\right|_{\text{phys}}=S_{\text{phys}}
\end{equation}
Let us introduce the following two BRST quartet of external sources
\begin{eqnarray}
s\bar{N}^{\alpha\beta}_{\mu\nu}=-\bar{M}^{\alpha\beta}_{\mu\nu}\,,\qquad 
s\bar{M}^{\alpha\beta}_{\mu\nu}=0\,,\qquad
sM^{\alpha\beta}_{\mu\nu}=N^{\alpha\beta}_{\mu\nu}\,,\qquad sN^{\alpha\beta}_{\mu\nu}=0&\nonumber\\
s\bar{U}^{\alpha\beta}=-\bar{V}^{\alpha\beta}\,,\qquad s\bar{V}^{\alpha\beta}=0\,,\qquad 
sV^{ab}=U^{\alpha\beta}\,,\qquad sU^{\alpha\beta}=0 &
\label{Sources_NMUV}
\end{eqnarray}
where $(\bar{N}^{\alpha\beta}_{\mu\nu},N^{\alpha\beta}_{\mu\nu}),(V^{\alpha\beta},\bar{V}^{\alpha\beta})$ are anti-commutating sources and  $(\bar{M}^{\alpha\beta}_{\mu\nu},M^{\alpha\beta}_{\mu\nu}),(U^{\alpha\beta},\bar{U}^{\alpha\beta})$ are commutating. The sources in \eqref{Sources_NMUV} are necessary to restore the broken BRST invariance of the model, which, as shown in \eqref{SoftBreak}, it is due to the presence of the parameters $\gamma^{2}$ and $\sigma^{2}$. Notice that this symmetry restoration is only possible because the breaking in \eqref{SoftBreak} is soft, namely it is proportional to a mass dimension two parameter \cite{Zwanziger:1992qr}. Thus, we can rewrite which contains the Gribov parameters $\gamma^2$ and $\sigma^2$ in \eqref{Sphys1} as a BRST-exact sources action in the following way
\begin{eqnarray}
S_{\text{inv}}&=&s\int d^4x\left\{-\bar{N}^{\alpha\beta}_{\mu\nu}\,\mathcal{D}^{\alpha\omega}_{\mu}\varphi^{\omega\beta}_{\nu}
+M^{\alpha\beta}_{\mu\nu}\,\mathcal{D}^{\alpha\omega}_{\mu}\bar\omega^{\omega\beta}_{\nu}
+g\varepsilon^{\alpha\beta}\left(-\bar{U}^{\alpha\omega}\phi\,\eta^{\omega\beta}+V^{\alpha\omega}\phi\,\bar{\theta}^{\omega\beta}\right)
\right.\nonumber\\
&&\left.-\chi\bar{N}^{\alpha\beta}_{\mu\nu}M^{\alpha\beta}_{\mu\nu}
-\bar{\chi}\bar{V}^{\alpha\beta}U^{\alpha\beta}
\right\}\nonumber\\
&=&\int d^4x\left\{\bar{M}^{\alpha\beta}_{\mu\nu}\,\mathcal{D}^{\alpha\gamma}_{\mu}\varphi^{\gamma\beta}_{\nu}
   +M^{\alpha\beta}_{\mu\nu}\left[\mathcal{D}^{\alpha\omega}_{\mu}\bar{\varphi}^{\omega\beta}_{\nu}+g\varepsilon^{\alpha\omega}
   \left(\partial_{\mu}c+g\varepsilon^{\rho\xi}A_{\mu}^\rho c^\xi\right)\bar{\omega}^{\omega\beta}_{\nu}\right]
   \right.\nonumber\\
&& +N^{\alpha\beta}_{\mu\nu}\,\mathcal{D}^{\alpha\omega}_{\mu}\bar{\omega}^{\omega\beta}_{\nu}
   +\bar{N}^{\alpha\beta}_{\mu\nu}\left[\mathcal{D}^{\alpha\omega}_{\mu}\bar{\omega}^{\omega\beta}_{\nu}+g\varepsilon^{\alpha\omega}
   \left(\partial_{\mu}c+g\varepsilon^{\rho\xi}A_{\mu}^\rho c^\xi\right)\varphi^{\omega\beta}_{\nu}\right]\nonumber\\
&&  +g\varepsilon^{\alpha\beta}\left(\bar{V}^{\alpha\omega}\phi\eta^{\omega\beta}+V^{\alpha\omega}(\phi\bar{\eta}^{\omega\beta}-g\varepsilon^{\rho\xi}\phi^\rho c^\xi\bar{\theta}^{\omega\beta})+U^{\alpha\omega}\phi\bar{\theta}^{\omega\beta}+\bar{U}^{\alpha\omega}(\phi\theta^{\omega\beta}-g\varepsilon^{\rho\xi}\phi^\rho c^\xi\eta^{\omega\beta})\right)
\nonumber\\
&&\left.+\chi\left(\bar{N}^{\alpha\beta}_{\mu\nu}N^{\alpha\beta}_{\mu\nu}+\bar{M}^{\alpha\beta}_{\mu\nu}M^{\alpha\beta}_{\mu\nu}\right)
+\bar{\chi}\left(\bar{V}^{\alpha\beta}V^{\alpha\beta}+\bar{U}^{\alpha\beta}U^{\alpha\beta}\right)
\right\}
\label{ActSUV}
\end{eqnarray}
%%%%%%%%%%%%%
The  original theory \eqref{Sphys1} is recovered by demanding that the sources attain a suitable physical limit, namely
\begin{eqnarray}
&-M^{\alpha\beta}_{\mu\nu}\Bigl|_{\text{phys}}=\bar{M}^{\alpha\beta}_{\mu\nu}\Bigl|_{\text{phys}}=\gamma^{2}\delta^{\alpha\beta}\delta^{\mu\nu}\,,\qquad
N^{\alpha\beta}_{\mu\nu}\Bigl|_{\mathrm{phys}}=\bar{N}^{\alpha\beta}_{\mu\nu}\Bigl|_{\mathrm{phys}}=0\,,\nonumber\\
&-V^{\alpha\beta}\Bigl|_{\mathrm{phys}}=\bar{V}^{\alpha\beta}\Bigl|_{\mathrm{phys}}=\sigma^{2}\delta^{\alpha\beta}\,,\qquad
U^{\alpha\beta}\Bigl|_{\mathrm{phys}}=\bar{U}^{\alpha\beta}\Bigl|_{\mathrm{phys}}=0
\label{PhysVal1}
\end{eqnarray}
%%%%%%%%%%%%%%%%%%%%%%%%%%%%%%%%%%%%%%%%%%%%%%%%%%%%%%%%%%%%%%%
We emphasize that one should notice that the presence of terms which are quadratic in the sources is allowed by power counting and has to be added for renormalizability purposes. The parameter $(\chi)$ stands for free coefficients related to the vacuum term in the gap equation (also called Zwanziger horizon condition)  which determined the Gribov parameter in a self-consistent way (see \cite{Capri:2010an}), however, $(\bar{\chi})$ doesn't have this geometric interpretation, since horizon-like matter term was motivated through for associate it with the effective dynamics of infrared, confined matter that we would like to obtain.  See \cite{Capri:2014bsa,Palhares:2016wqn} for a more detailed discussion on this topic.

We call a particular attention to the new enlarged BRST-exact action, given by
\begin{equation}
\Sigma_{0}=S_{\text{YM}}+S_{\text{MAG}}+S_{\varphi\omega}+S_{\text{matter}}+S_{\eta\theta}+S_{\text{inv}} 
\label{Sigma_0}
\end{equation}
 displays two new global symmetries, a  U(8)-symmetry for pure gauge sector and a U(2)-symmetry for the matter sector, meaning that we can make use of the composite indixes, or multi-indixes notation, $I,J,K,\dots\equiv\{\alpha,\mu\}=1,\dots, 8$ and the indixes $i,j,k,\dots\equiv \alpha=1,2$,  to representing global $U(8)$ and $U(2)$ symmetric combinations, respectively. The last multi-index is not really a composite one, but the usual off-diagonal color index, and they are introduced just to standardize the notation. The indixes $I,J,K,\dots$ provide that contractions like $\bar{\varphi}^{\alpha\beta}_{\mu}\varphi^{\alpha\beta}_{\mu}$ could be written as $\bar{\varphi}^{\alpha}_{I}\varphi^{\alpha}_{I}$ and forbid contractions like $A^{\alpha}_{\mu}A^{\beta}_{\nu}\partial_{\nu}\varphi^{\alpha\beta}_{\mu}$. In fact, the $U(8)$-symmetry provides that  only the Zwanziger fields and its related sources can contract with the multi-indixes, then, one can write that
\begin{eqnarray}
\left(\bar{\varphi}^{\alpha\beta}_{\mu},\varphi^{\alpha\beta}_{\mu},\bar{\omega}^{\alpha\beta}_{\mu},\omega^{\alpha\beta}_{\mu}\right)&\equiv &
\left(\bar{\varphi}^{\alpha} _I,\varphi^{\alpha}_I,\bar{\omega}^{\alpha}_I,\omega^a_I\right)
\nonumber\\
\left(\bar{M}^{\alpha\beta}_{\mu\nu},M^{\alpha\beta}_{\mu\nu},\bar{N}^{\alpha\beta}_{\mu\nu},N^{\alpha\beta}_{\mu\nu}\right)&\equiv&
\left(\bar{M}^{\alpha}_{\mu I},M^{\alpha}_{\mu I},\bar{N}^{\alpha}_{\mu I},N^{\alpha}_{\mu I}\right)
\end{eqnarray}
The $U(8)$-symmetry is broken when the sources attain their physical values \eqref{PhysVal1}. Analogously, the $U(2)$ symmetry provides special contractions among the localizing matter sector fields and its related sources. For instance, contractions like $\bar\eta^{\alpha\beta}\eta^{\alpha\beta}\equiv\bar\eta^{\alpha}_{i}\eta^{\alpha}_{i}$ are allowed while contractions like $\phi^{\alpha}A^{\beta}_{\mu}\partial_{\mu}\eta^{\alpha\beta}$ or $\varphi^{\alpha\beta}_{\mu}\partial_{\mu}\eta^{\alpha\beta}$ are forbidden. Thus one can write that
\begin{eqnarray}
\left(\bar{\eta}^{\alpha\beta},\eta^{\alpha\beta},\bar{\theta}^{\alpha\beta},\theta^{ab}\right) &\equiv&
\left(\bar{\eta}^{\alpha}_i,\eta^{\alpha}_i,\bar{\theta}^{\alpha}_i,\theta^{\alpha}_i\right)
\nonumber\\
\left(\bar{U}^{\alpha\beta},U^{\alpha\beta},\bar{V}^{\alpha\beta},V^{\alpha\beta}\right)&\equiv&
\left(\bar{U}^{\alpha}_i,U^{\alpha}_i,\bar{V}^{\alpha}_i,V^{\alpha}_i\right)
\end{eqnarray}
The $U(2)$ symmetry is also broken in the physical limit \eqref{PhysVal1}. These two global symmetries can be expressed in a functional form by\footnote{See \eqref{WI_QIJ} and \eqref{WI_Qij} for the definition of both the operators.}
\begin{equation}
Q_{IJ}\left(\Sigma_{0}\right)=0\,,\qquad
Q_{ij}\left(\Sigma_{0}\right)=0
\end{equation}
in such a way that the trace of the operators $Q_{IJ}$ and $Q_{ij}$ define new additional conserved quantum number in the auxiliary localizing Zwanziger(-like) sector, the $Q_8$ and $Q_2$ charge. The corresponding values of these charges for each field and source are summarized in the Tables \ref{Tab:gaugeSector}, \ref{Tab:matterSector} and \ref{Tab:ExtSoruces} in the end of the present  section.

Besides that, we notice that the BRST transformations of the gauge, ghost and matter fields in \eqref{BRSTscalar}  are  non-linear. In implementation of algebraic renormalization procedure, we need to properly take into  account the corresponding composite operators, a task which is achieved by introducing a set  of external sources coupled to the non-linear BRST transformations
\begin{eqnarray}
\Sigma_{\text{ext}}^{(1)}&=&s\int d^4x\left\{-\Omega_{\mu}^{\alpha}A_{\mu}^{\alpha}
-g\varepsilon^{\alpha\beta}\xi_{\mu}^{\alpha}A_{\mu}^\beta c
-\Omega_{\mu}A_{\mu}+L^{\alpha}c^{\alpha} + Lc-\tilde\Omega^{\alpha}\phi^{\alpha}-g\varepsilon^{\alpha\beta}\tilde\xi^{\alpha}\phi^\beta c
-\tilde\Omega\phi\right\}
\nonumber\\
&=&\int d^4x\left\{-\Omega_{\mu}^\alpha \mathcal{D}_{\mu}^{\alpha\beta}c^\beta 
-g\varepsilon^{\alpha\beta}\tau^\alpha_{\mu} A_{\mu}^\beta c
+\xi_{\mu}^\alpha \left[g\varepsilon^{\alpha\beta} \left(\mathcal{D}_{\mu}^{\beta\omega} c^\omega\right)c-
\frac{g^2}{2}\varepsilon^{\alpha\beta}\varepsilon^{\omega\rho}A_{\mu}^\beta c^\omega c^\rho \right]
\right.\nonumber\\
&& -\Omega_{\mu}\left(\partial_{\mu}c+g\varepsilon^{\alpha\beta}A_{\mu}^\alpha c^\beta\right)
+g\varepsilon^{\alpha\beta}L^\alpha c^\beta c+\frac{g^2}{2}\varepsilon^{\alpha\beta}Lc^\alpha c^\beta
+\tilde\Omega^\alpha \varepsilon^{\alpha\beta}\phi c^\beta
-g\varepsilon^{\alpha\beta}\tilde\tau^\alpha \phi^\beta c\nonumber\\
&&\left.-g^2\varepsilon^{\alpha\beta}\tilde\xi^\alpha \left(\varepsilon^{\beta\omega}\phi\,c^\omega c
+\frac{1}{2}\varepsilon^{\omega\rho}\phi^\beta c^\omega c^\rho\right)
-\tilde\Omega g\varepsilon^{\alpha\beta}\phi^\alpha c^\beta
\right\}
\label{ExtSoruces01}
\end{eqnarray}
where to guarantee the BRST invariance we require that all sources are $s-$invariants except for
\begin{equation}
s\xi_{\mu}^{\alpha}=-(\Omega_{\mu}^{\alpha}-\tau_{\mu}^{\alpha})\,,\qquad s\tilde{\xi}_{\mu}^{\alpha}=-(\tilde{\Omega}_{\mu}^{\alpha}-\tilde{\tau}_{\mu}^{\alpha})
\label{ExtSoruces02}
\end{equation} 
and thus providing the Slavnov-Taylor identity. After the renormalization procedure they can be taken to zero. Furthermore, as was shown in \cite{Capri:2006cz}, the gauge sector in the extended action ($\Sigma_0$ term in the  equation \eqref{Sigma_0} plus the external sources in \eqref{ExtSoruces01}) displays a rather rich symmetry content. In fact, there exist several additional  symmetries involving the exchange between the Faddeev-Popov fields and the localizing auxiliary fields, namely symmetries which mix the gauge fixing and functional space restriction sectors. Such a set of transformations are given by \cite{Capri:2006cz}
\begin{itemize}
\item{$\delta_I$-symmetry
\begin{equation}
\delta_I\bar{c}^\alpha=\varphi^\alpha_I\,,\quad
\delta_I\bar{\varphi}^\alpha_J=\delta_{IJ}c^\alpha\,,\quad
\delta_Ib^\alpha=g\varepsilon^{\alpha\beta}\varphi^\beta_Ic\,,\quad
\delta_I \Omega^\alpha_{\mu}=M^\alpha_{\mu I}
\label{delta1gauge}
\end{equation}
}
\item{$\bar{\delta}_I$-symmetry
\begin{equation}
\bar{\delta}_I\bar{c}^\alpha=\bar{\omega}^\alpha_I\,,\quad
\bar{\delta}_I\omega^\alpha_J=-\delta_{IJ}c^\alpha\,,\quad
\bar{\delta}_Ib^\alpha=g\varepsilon^{\alpha\beta}\bar{\omega}^\beta_Ic\,,\quad
\bar{\delta}_I\Omega^\alpha_{\mu}=-\bar{N}^\alpha_I
\label{delta2gauge}
\end{equation}
}
\item{$d_I$-symmetry
\begin{eqnarray}
d_I\bar{c}^\alpha=\omega^\alpha_I+g\varepsilon^{\alpha\beta}\varphi^\beta_I c\,,\quad
d_I\bar{\varphi}^\alpha_J=\delta_{IJ}g\varepsilon^{\alpha\beta}c^\beta c\,,\quad
d_Ib^\alpha=g\varepsilon^{\alpha\beta}\omega^\beta_I c
+\frac{g^2}{2}\varepsilon^{\alpha\beta}\varepsilon^{\omega\rho}\varphi_I^\beta c^\omega c^\rho\,, \nonumber\\
d_I\bar{\omega}^\alpha_J=\delta_{IJ}c^\alpha\,,\quad
d_I\Omega^\alpha_{\mu}=N^\alpha_{\mu I}\,,\quad
d_I\xi^\alpha_{\mu}=-M^\alpha_{\mu I}\qquad\qquad
\label{delta3gauge}
\end{eqnarray}
}
\item{$\bar{d}_I$-symmetry
\begin{eqnarray}
\bar{d}_I\bar{c}^\alpha=-\varphi^\alpha_I+g\varepsilon^{\alpha\beta}\bar{\omega}^\beta_I c\,,\quad
\bar{d}_I\omega^\alpha_J=\delta_{IJ}g\varepsilon^{\alpha\beta}c^\beta c\,,\quad
\bar{d}_Ib^\alpha =-g\varepsilon^{\alpha\beta}\bar{\varphi}^\beta_I c
+\frac{g^2}{2}\varepsilon^{\alpha\beta}\varepsilon^{\omega\rho}\bar{\omega}_I^\beta c^\omega c^\rho\,, \nonumber\\
\bar{d}_I\varphi^\alpha_J=-\delta_{IJ}c^\alpha\,,\quad
\bar{d}_I\Omega^\alpha_{\mu}=-\bar{M}^\alpha_{\mu I}\,,\quad
\bar{d}_I\xi^\alpha_{\mu}=\bar{N}^\alpha_{\mu I}\qquad\qquad
\label{delta4gauge}
\end{eqnarray}
}
\end{itemize}
These symmetries are very important to guaranteed the perturbative renormalizability of the model, since they allow us to keep under control a new class of quartic terms  allowed by power counting which arise due to the nonlinearity of the gauge condition \cite{Capri:2006cz}.  Furthermore, in \cite{Capri:2008ak} was shown that, in addition to quartic terms in Nakanishi-Lautrup and ghost fields \cite{Min:1985bx,Fazio:2001rm,Gracey:2005vu}, a new set of quartic terms in Zwanziger fields are necessary to taking into account a new class of UV-divergent Feynman diagrams when the horizon effects are relevant due to the non-trivial dynamics of the fields $(\phi,\bar{\phi},\omega,\bar{\omega})$. The $\delta-$ and $d-$ symmetries require  these quartic terms to be proportional to a unique free parameter, namely, the gauge parameter. As a consequence, the limit $\alpha\to 0$, which allow us to recover the MAG condition, implies in the vanishing of the quartic terms in the physical limit.

A remarkable fact is the existence of a new set of symmetries for auxiliary fields in the matter sector, in perfect analogy to the symmetries above, which are given by:
\begin{itemize}
\item{$\delta_i$-symmetry
\begin{equation}
\delta_i\bar{c}^\alpha=\eta^\alpha_i\,,\quad\delta_i\bar{\eta}^\alpha_j=\delta_{ij}c^\alpha\,,\quad
\delta_ib^\alpha=g\varepsilon^{\alpha\beta}\eta^\beta_i c\,,\quad
\delta_i\tilde{\Omega}^\alpha=V^\alpha_i
\label{delta1scalar}
\end{equation}
}
\item{$\bar{\delta}_i$-symmetry
\begin{equation}
\bar{\delta}_i\bar{c}^\alpha=\bar{\theta}^\alpha_i\,,\quad
\bar{\delta}_i\theta^\alpha_j=-\delta_{ij}c^\alpha\,,\quad
\bar{\delta}_ib^\alpha=g\varepsilon^{\alpha\beta}\bar{\theta}^\beta_i c\,,\quad
\bar{\delta}_i\tilde{\Omega}^\alpha=-U^\alpha_i
\label{delta2scalar}
\end{equation}
}
\item{$d_i$-symmetry
\begin{eqnarray}
d_i\bar{c}^\alpha=\theta^\alpha_i+g\varepsilon^{\alpha\beta}\eta^\beta_ic\,,\quad
d_i\bar{\eta}^\alpha_j=\delta_{ij}g\varepsilon^{\alpha\beta}c^\beta c\,,\quad
d_ib^\alpha=g\varepsilon^{\alpha\beta}\theta^\beta_ic
+\frac{g^2}{2}\varepsilon^{\alpha\beta}\varepsilon^{\omega\rho}\eta_i^\beta c^\omega c^\rho\,, \nonumber\\
d_i\bar{\theta}^\alpha_j=\delta_{ij}c^\alpha\,,\quad
d_i\tilde{\Omega}^\alpha=U^\alpha_i\,,\quad
d_i\tilde{\xi}^\alpha=-V^\alpha_i\qquad\qquad
\label{delta3scalar}
\end{eqnarray}
}
\item{$\bar{d}_i$-symmetry
\begin{eqnarray}
\bar{d}_i\bar{c}^\alpha=-\eta^\alpha_i+g\varepsilon^{\alpha\beta}\bar{\theta}^\beta_i c\,,\quad
\bar{d}_i\theta^\alpha_j=\delta_{ij}g\varepsilon^{\alpha\beta}c^\beta c\,,\quad
\bar{d}_ib^\alpha=-g\varepsilon^{\alpha\beta}\bar{\eta}^b_ic
+\frac{g^2}{2}\varepsilon^{\alpha\beta}\varepsilon^{\omega\rho}\bar{\theta}_i^\beta c^\omega c^\rho\,, \nonumber\\
\bar{d}_i\eta^\alpha_j=-\delta_{ij}c^\alpha\,,\quad
\bar{d}_i\tilde{\Omega}^\alpha=-\bar{V}^\alpha_i\,,\quad
\bar{d}_i\tilde{\xi}^\alpha=\bar{U}^\alpha_i\qquad\qquad
\label{delta4scalar}
\end{eqnarray}
}
\end{itemize}
%%%%%%%%%%%%%%%%%%%%%%%%%%%%%%%%%%%%%%%%%%%%%
%%%%%%%%%%%%%%%%%%%%%%%%%%%%%%%%%%%%%%%%%%%%%
%   In particular, these symmetries  keep under control the number of new quartic terms allowed by power counting.
%%%%%%%%%%%%%%%%%%%%%%%%%%%%%%%%%%%%%%%%%%%%%
This will ensure the renormalizability of the new Zwanziger-like matter field sector, in the same way as for the gauge sector. On the other hand, the inclusion of interacting scalar fields  generate a new class of UV-divergent Feynman diagramas, as showed in a previus work \cite{Capri:2015pxa}. Hence, we need to add another quartic terms in a BRST invariant fashion in order to  renormalizes these new divergences, which, like in pure gauge sector, must be generalized to  the localizing Zwanziger fields of the horizon function of matter sector too, in such a way  that satisfies all original symmetries of  gauge sector and the new $\tilde{\delta}-$ and $\tilde{d}-$symmetries above.  After a simple algebra, we obtain that the most general term necessary to deal with the new divergences which obey the full set Ward identities displays in the next section, is given by 
\begin{eqnarray}
\Sigma_{\text{qua}}&=&\alpha\,s\int d^4x\left\{
\bar{c}^\alpha b^\alpha-g\varepsilon^{\alpha\beta}\bar{c}^\alpha\bar{c}^\beta c
-2g^2\bar{c}^\alpha c^\alpha\left(\varphi^\beta_I\bar{\omega}^\beta_I+\eta^\beta_i\bar{\theta}^\beta_i\right)
+2g^2\bar{\omega}^\alpha_I\varphi^\alpha_I\left(\bar{\varphi}^\beta_J\varphi^\beta_J-\bar{\omega}^\beta_J\omega^\beta_J\right)
\right.\nonumber\\
&&\left.
+2g^2\bar{\theta}^\alpha_i\eta^\alpha_i\left(\bar{\eta}^\beta_j\eta^\beta_j-\bar{\theta}^\beta_j\theta^\beta_j\right)
+g^2\varphi^\alpha_I\bar{\omega}^\alpha_I\left(\bar{\eta}^\beta_j\eta^\beta_j-\bar{\theta}^\beta_j\theta^\beta_j\right)
+g^2\eta^\alpha_i\bar{\theta}^\alpha_i\left(\bar{\varphi}^\beta_J\varphi^\beta_J-\bar{\omega}^\beta_J\omega^\beta_J\right)
\right\} \nonumber\\
&&+\beta\,s\int d^4x\left\{\varepsilon^{\alpha\beta}\phi\phi^\alpha\bar{c}^\beta
+g\phi^\alpha\phi^\alpha\left(\varphi^\beta_I\bar{\omega}^\beta_I+\eta^\beta_i\bar{\theta}^\beta_i\right)
-g\phi^\alpha\phi^\beta\left(\varphi^\beta_I\bar{\omega}^\alpha_I+\eta^\beta_i\bar{\theta}^\alpha_i\right)
-g\phi^2\left(\eta^\beta_i\bar{\theta}^\beta_i-\varphi^\beta_I\bar{\omega}^\beta_I\right)
\right\}\nonumber\\
&=&\alpha\int d^4x\left\{b^\alpha b^\alpha -2g\varepsilon^{\alpha\beta}b^\alpha\bar{c}^\beta c
+g^2\bar{c}^\alpha\bar{c}^\beta c^\alpha c^\beta -2g^2\left(b^\alpha c^\alpha -g\varepsilon^{\alpha\omega}\bar{c}^\alpha c^\omega c\right)
\left(\varphi^\beta_I\bar{\omega}^\beta_I+\eta^\beta_i\bar{\theta}^\beta_i\right)\right.\nonumber\\
&&
-2g^2\bar{c}^\alpha c^\alpha\left(\bar{\varphi}^\beta_I\varphi^\beta_I-\bar{\omega}^\beta_I\omega^\beta_I
+\bar{\eta}^\beta_i\eta^\beta_i-\bar{\theta}^\beta_i\theta^\beta_i\right)
+2g^2\left(\bar{\varphi}^\alpha_I\varphi^\alpha_I-\bar{\omega}^\alpha_I\omega^\alpha_I\right)
\left(\bar{\eta}^\beta_i\eta^\beta_i-\bar{\theta}^\beta_i\theta^\beta_i\right)\nonumber\\
&&
\left.+2g^2\left(\bar{\varphi}^\alpha_I\varphi^\alpha_I-\bar{\omega}^\alpha_I\omega^\alpha_I\right)
\left(\bar{\varphi}^\beta_J\varphi^\beta_J-\bar{\omega}^\beta_J\omega^\beta_J\right)
+2g^2\left(\bar{\eta}^\alpha_i\eta^\alpha_i-\bar{\theta}^\alpha_i\theta^\alpha_i\right)
\left(\bar{\eta}^\beta_j\eta^\beta_j-\bar{\theta}^\beta_j\theta^\beta_j\right)\right\}\nonumber\\
&&
+\beta\int d^4x\left\{g\phi^\alpha\phi^\alpha\left(\bar{\varphi}^\beta_I\varphi^\beta_I-\bar{\omega}^\beta_I\omega^\beta_I
+\bar{\eta}^\beta_i\eta^\beta_i-\bar{\theta}^\beta_i\theta^\beta_i-c^\beta\bar{c}^\beta\right)\right.\nonumber\\
&&
+g\phi^\alpha\phi^\beta\left(\bar{\varphi}^\alpha_I\varphi^\beta_I-\bar{\omega}^\alpha_I\omega^\beta_I
+\bar{\eta}^\alpha_i\eta^\beta_i-\bar{\theta}^\alpha_i\theta^\beta_i+c^\alpha\bar{c}^\beta\right)
+\phi\phi^\alpha\left(gc\bar{c}^\alpha +\varepsilon^{\alpha\beta}b^\beta\right) \nonumber\\
&&
+g\phi\phi\left(\bar{\varphi}^\alpha_I\varphi^\alpha_I-\bar{\omega}^\alpha_I\omega^\alpha_I
-\bar{\eta}^\alpha_i\eta^\alpha_i+\bar{\theta}^\alpha_i\theta^\alpha_i+c^\alpha\bar{c}^\alpha\right)
-g^2\phi^\alpha\left(\varepsilon^{\alpha\rho}\phi^\omega c+\varepsilon^{\alpha\omega}\phi^\rho c\right)
\left(\varphi^\omega_I\bar{\omega}^\rho_I+\eta^\omega_i\bar{\theta}^\rho_i\right)\nonumber\\
&& \left. 
+g^2\phi\phi^\alpha\left(\varepsilon^{\alpha\rho}c^\omega+\varepsilon^{\alpha\omega}c^\rho\right)
\left(\eta^\omega_i\bar{\theta}^\rho_i-\varphi^\omega_I\bar{\omega}^\rho_I\right)
+2g^2\varepsilon^{\omega\rho}\phi^\omega c^\rho
\left(\eta^\alpha_i\bar{\theta}^\alpha_i\varphi^\alpha_I\bar{\omega}^\alpha_I\right)\right\}
\label{Sigma_quartic}
\end{eqnarray}
where $\beta$ stands for a second gauge-like parameter, which was introduced in \cite{Capri:2015pxa}\footnote{See the mentioned reference for a detailed discussion on this subject.}.  All the terms in \eqref{Sigma_quartic} are proportional to the parameters $\alpha$ or $\beta$, therefore they are associated to nonlinearity of the off-diagonal gauge condition, $\partial_{\mu}A^{\alpha}_{\mu}=g\varepsilon^{\alpha\beta}A_{\mu}A^{\beta}_{\mu}$. Such nonlinearity generates extra interaction vertices\footnote{When compared to the Landau and linear covariant gauges.} that give rise not only to quartic ghost interaction terms, but ghost-Zwanziger localizing fields  interaction terms too. Explicitly, in the present case, these extra interaction terms are generalized to include the localizing auxiliary fields of the scalar matter sector $\{\eta,\bar\eta,\theta,\bar\theta\}$ in such a way to preserve the identities (\ref{delta1scalar} -- \ref{delta4scalar}). Looking in the equation of motion for the off-diagonal Nakanishi-Lautrup field
\begin{equation}
\frac{\delta S}{\delta b^\alpha}=\mathcal{D}_{\mu}^{\alpha\beta}A_{\mu}^\beta+\alpha\left(b^\alpha-g\varepsilon^{\alpha\beta}\bar{c}^\beta
-2g^2c^\alpha(\varphi^\beta_I\bar{\omega}^\beta_I+\eta^\beta_i\bar{\theta}^\beta_i)\right)
+\frac{\beta}{2}g\varepsilon^{\alpha\beta}\phi\phi^\beta
\end{equation}
we can perceive that the original maximal Abelian gauge condition is recovered in the limit $\alpha,\beta\rightarrow 0$. New composite operators are contained in these symmetries, thus we need to add two new sets of BRST doublets of external sources $(X^\alpha_I,Y^\alpha_I)$ and $(\bar{X}^\alpha_I,\bar{Y}^\alpha_I)$  for gauge sector, and two BRST doublets for matter sector, $(\tilde{X}^\alpha_i,\tilde{Y}^\alpha_i)$ and $(\bar{\tilde{X}}^\alpha,_i\bar{\tilde{Y}}^\alpha_i)$, so that
\begin{eqnarray}
\Sigma_{\text{ext}}^{(2)}&=&s\int d^4x\,g\varepsilon^{\alpha\beta}\left(\bar{X}^\alpha_{I}\varphi^\beta_I c
+Y^\alpha_I\bar{\omega}^\beta_I c+\bar{\tilde{X}}^\alpha_i\eta^\beta_i c
+\tilde{Y}^\alpha_i\bar{\theta}^\beta_i c\right)\nonumber\\
&=&g\varepsilon^{\alpha\beta}\int d^4x\,\left\{\bar{Y}^\alpha_I\varphi^\beta_I c
+X^\alpha_I\bar{\omega}^\beta_Ic
+\bar{X}^\alpha_I\left(\omega^\beta_Ic
+\frac{g}{2}\varepsilon^{\omega\rho}\varphi^\beta_I c^\omega c^\rho\right)
-Y^\alpha_I\left(\bar{\varphi}^\beta_I c
-\frac{g}{2}\varepsilon^{\omega\rho}\bar{\omega}^\beta_I c^\omega c^\rho\right)
\right.\nonumber\\
&&\left. +\bar{\tilde{X}}^\alpha_i\left(\theta^\beta_ic
+\frac{g}{2}\varepsilon^{\omega\rho}\eta^\beta_ic^\omega c^\rho\right)
+\tilde{X}^\alpha_i\bar{\theta}^\beta_i c
+\bar{\tilde{Y}}^\alpha_i\eta^\beta_i c
-\tilde{Y}^\alpha_i\left(\bar{\eta}^\beta_i c
-\frac{g}{2}\varepsilon^{\omega\rho}\bar{\theta}^\beta_i c^\omega c^\rho\right)\right\}
\label{Sigma_ext2}
\end{eqnarray}
where the $s-$transformations for the new set of external sources for composite operators is
\begin{eqnarray}
sY_I^\alpha=X^\alpha_I\,,\qquad X_I^\alpha=0\,,\qquad 
s\bar{X}_I^\alpha=-\bar{Y}^\alpha_I\,,\qquad \bar{Y}_I^\alpha=0\,;&\nonumber\\
s\tilde{Y}_i^\alpha=\tilde{X}^\alpha_i\,,\qquad \tilde{X}_i^\alpha=0\,,\qquad 
s\bar{\tilde{X}}_i^\alpha=-\bar{\tilde{Y}}^\alpha_i\,,\qquad \bar{\tilde{Y}}_i^\alpha=0 &
\end{eqnarray}
%%%%%%%%%%%%%%%%%%%%%4

We emphasize that the sources in \eqref{Sigma_ext2} have the same role of the BRST external sources, but in this case, they are necessary to take into account nonlinear transformations (\ref{delta1gauge} -- \ref{delta4scalar}). Quantum numbers for all fields and sources are display in the Tables 1, 2 and 3. Thus, finally, the full local and BRST invariant action for Yang-Mills theory with scalar matter which implements non-perturbative effects \textit{\`a la Gribov-Zwanziger}, $\Sigma$, is given by
\begin{eqnarray}
\Sigma&=&\int d^4x\left\{\frac{1}{4}\left(F_{\mu\nu}^\alpha F_{\mu\nu}^\alpha +F_{\mu\nu}F_{\mu\nu}\right)
+b^\alpha \mathcal{D}_{\mu}^{\alpha\beta}A_{\mu}^\beta-\bar{c}^\alpha\mathcal{M}^{\alpha\beta}c^\beta
+g\varepsilon^{\alpha\beta}\bar{c}^\alpha c\, \mathcal{D}_{\mu}^{\beta\omega}A_{\mu}^\omega
+b\partial_{\mu}A_{\mu}
\right.\nonumber\cr
&& 
+\bar{c}\,\partial_{\mu}\left(\partial_{\mu}c+g\varepsilon^{\alpha\beta}A_{\mu}^\alpha c^\beta \right)
+\bar{\varphi}^\alpha_{I}\mathcal{M}^{\alpha\beta} \varphi^{\beta}_{I}
-\bar{\omega}^\alpha_I\mathcal{M}^{\alpha\beta}\omega^{\beta}_I
+\bar{\omega}^\alpha_I\mathcal{F}^{\alpha\beta}\varphi^{\beta}_I
+\bar{M}^\alpha_{\mu I}\,\mathcal{D}^{\alpha\beta}_{\mu}\varphi^{\beta}_I
\nonumber\cr
&&
+M^\alpha_{\mu I}\left[\mathcal{D}^{\alpha\beta}_{\mu}\bar{\varphi}^\beta_I
+g\varepsilon^{\alpha\beta}\left(\partial_{\mu}c+g\varepsilon^{\omega\rho}A_{\mu}^\omega c^\rho\right)\bar{\omega}^\beta_I\right]
 +N^\alpha_{\mu I}\,\mathcal{D}^{\alpha\beta}_{\mu}\bar{\omega}^\beta_I
\nonumber\cr
&&
   +\bar{N}^\alpha_{\mu\nu}\left[D^{\alpha\beta}_{\mu}\bar{\omega}^\beta_I+g\varepsilon^{\alpha\beta}
   \left(\partial_{\mu}c+g\varepsilon^{\omega\rho}A_{\mu}^\omega c^\rho\right)\varphi^\beta_I\right]
   +\chi(\bar{M}_{\mu I}^\alpha M_{\mu I}^\alpha+\bar{N}_{\mu I}^\alpha N_{\mu I}^\alpha) \nonumber\cr
&&
+(\partial_{\mu}\phi^\alpha)(\partial_{\mu}\phi^\alpha)+(\partial_{\mu}\phi)(\partial_{\mu}\phi)
-2g^2\varepsilon^{\alpha\beta}\left[(\partial_{\mu}\phi)\phi^\alpha A_{\mu}^\beta
-(\partial_{\mu}\phi^\alpha)\phi A_{\mu}^\beta+(\partial_{\mu}\phi^\alpha)\phi^\beta A_{\mu}\right]\nonumber\cr
&&
+g^2\left[A_{\mu}^\alpha A_{\mu}^\alpha\left(\phi^\beta\phi^\beta +\phi\phi\right)
+A_{\mu}A_{\mu}\phi^\alpha \phi^\alpha -A_{\mu}^\alpha A_{\mu}^\beta \phi^\alpha \phi^\beta
-2A_{\mu}^\alpha A_{\mu}\phi^\alpha\phi\right]
\nonumber\cr
&&
+\frac{m^2_{\phi}}{2}\left(\phi^\alpha \phi^\alpha+\phi\phi\right)
+\frac{\lambda}{4!}\left[\left(\phi^\alpha\phi^\alpha\right)^2+2\phi^2\phi^\alpha \phi^\alpha+\phi^4\right]
+\bar{\eta}^{\alpha\beta}\mathcal{M}^{\alpha\omega}\eta^{\omega\beta}
\nonumber\cr
&&%%%%%%%%%%%%%%%%%%%
-\bar{\theta}^{\alpha\beta}\mathcal{M}^{\alpha\omega}\theta^{\omega\beta}
+\bar{\theta}^{\alpha\beta}\mathcal{F}^{\alpha\omega}\eta^{\omega\beta}
+\tilde{\chi}(\bar{V}_i^\alpha V_i^\alpha +\bar{U}_i^\alpha U_i^\alpha)\nonumber\cr
&&
+g\varepsilon^{\alpha\beta}\left(\bar{V}^{\alpha\omega}\phi\eta^{\omega\beta}+V^{\alpha\omega}(\phi\bar{\eta}^{\omega\beta}-g\varepsilon^{\rho\xi}\phi^\rho c^\xi\bar{\theta}^{\omega\beta})+U^{\alpha\omega}\phi\bar{\theta}^{\omega\beta}+\bar{U}^{\alpha\omega}(\phi\theta^{\omega\beta}-g\varepsilon^{\rho\xi}\phi^\rho c^\xi\eta^{\omega\beta})\right)\nonumber\cr
&&%%%%%%%%%%%%%%%%%%%
+\alpha\left[b^\alpha b^\alpha-2g\varepsilon^{\alpha\beta}b^\alpha\bar{c}^\beta c
+g^2\bar{c}^\alpha \bar{c}^\beta c^\alpha c^\beta -2g^2\left(b^\alpha c^\alpha-g\varepsilon^{\alpha\omega}\bar{c}^\alpha c^\omega c\right)
\left(\varphi^\beta_I\bar{\omega}^\beta_I+\eta^b_i\bar{\theta}^\beta_i\right)\right.\nonumber\cr
&&
-2g^2\bar{c}^\alpha c^\alpha\left(\bar{\varphi}^\beta_I\varphi^\beta_I-\bar{\omega}^\beta_I\omega^b_I
+\bar{\eta}^\beta_i\eta^\beta_i-\bar{\theta}^\beta_i\theta^\beta_i\right)
+2g^2\left(\bar{\varphi}^\alpha_I\varphi^\alpha_I-\bar{\omega}^\alpha_I\omega^\alpha_I\right)
\left(\bar{\eta}^\beta_i\eta^\beta_i-\bar{\theta}^\beta_i\theta^\beta_i\right)\nonumber\cr
&&
\left.+2g^2\left(\bar{\varphi}^\alpha_I\varphi^\alpha_I-\bar{\omega}^\alpha_I\omega^\alpha_I\right)
\left(\bar{\varphi}^\beta_J\varphi^\beta_J-\bar{\omega}^\beta_J\omega^\beta_J\right)
+2g^2\left(\bar{\eta}^\alpha_i\eta^\alpha_i-\bar{\theta}^\alpha_i\theta^\alpha_i\right)
\left(\bar{\eta}^\beta_j\eta^\beta_j-\bar{\theta}^\beta_j\theta^\beta_j\right)\right]\nonumber\cr
&&
+\beta\left[g\phi^\alpha\phi^\alpha\left(\bar{\varphi}^\beta_I\varphi^\beta_I-\bar{\omega}^\beta_I\omega^\beta_I
+\bar{\eta}^\beta_i\eta^\beta_i-\bar{\theta}^\beta_i\theta^\beta_i-c^\beta\bar{c}^\beta\right)\right.\nonumber\cr
&&
+g\phi^\alpha\phi^\beta\left(\bar{\varphi}^\alpha_I\varphi^\beta_I-\bar{\omega}^\alpha_I\omega^\beta_I
+\bar{\eta}^\alpha_i\eta^\beta_i-\bar{\theta}^\alpha_i\theta^\beta_i+c^\alpha\bar{c}^\beta\right)
+\phi\phi^\alpha\left(gc\bar{c}^\alpha+\varepsilon^{\alpha\beta}b^\beta\right) \nonumber\cr
&&
+g\phi\phi\left(\bar{\varphi}^\alpha_I\varphi^\alpha_I-\bar{\omega}^\alpha_I\omega^\alpha_I
-\bar{\eta}^\alpha_i\eta^\alpha_i+\bar{\theta}^\alpha_i\theta^\alpha_i+c^\alpha\bar{c}^\alpha\right)
-g^2\phi^\alpha\left(\varepsilon^{\alpha\rho}\phi^\omega c+\varepsilon^{\alpha\omega}\phi^\rho c\right)
\left(\varphi^\omega_I\bar{\omega}^\rho_I+\eta^\omega_i\bar{\theta}^\rho_i\right)\nonumber\cr
&&
\left.+g^2\phi\phi^\alpha\left(\varepsilon^{\alpha\rho}c^\omega +\varepsilon^{\alpha\omega}c^\rho\right)
\left(\eta^\omega_i\bar{\theta}^\rho_i-\varphi^\omega_I\bar{\omega}^\rho_I\right)
+2g^2\varepsilon^{\omega\rho}\phi^\omega c^\rho
\left(\eta^\alpha_i\bar{\theta}^\alpha_i + \varphi^\alpha_I \bar{\omega}^\alpha_I\right)\right]\nonumber\cr
&&
-\Omega_{\mu}^\alpha D_{\mu}^{\alpha\beta}c^\beta -g\varepsilon^{\alpha\beta}\tau^\alpha_{\mu}A_{\mu}^\beta c
+\xi_{\mu}^\alpha\left[g\varepsilon^{\alpha\beta}\left(D_{\mu}^{\beta\omega}c^\omega\right)c-
\frac{g^2}{2}\varepsilon^{\alpha\beta}\varepsilon^{\omega\rho}A_{\mu}^\beta c^\omega c^\rho\right]\nonumber\cr
&&
-\Omega_{\mu}\left(\partial_{\mu}c+g\varepsilon^{\alpha\beta}A_{\mu}^\alpha c^\beta\right)
+g\varepsilon^{\alpha\beta}L^\alpha c^\beta c+\frac{g^2}{2}\varepsilon^{\alpha\beta}Lc^\alpha c^\beta
+g\varepsilon^{\alpha\beta}\bar{Y}^{\alpha\omega}_{\mu}\varphi^{\beta\omega}_{\mu}c
+g\varepsilon^{\alpha\beta}X^{\alpha\omega}_{\mu}\bar{\omega}^{\beta\omega}_{\mu}c \nonumber\cr
&&
+g\varepsilon^{\alpha\beta}\bar{X}^{\alpha\omega}_{\mu}\left(\omega^{\beta\omega}_{\mu}c
+\frac{g}{2}\varepsilon^{\rho\xi}\varphi^{\beta\omega}_{\mu}c^\rho c^\xi\right)
-g\varepsilon^{\alpha\beta}Y^{\alpha\omega}_{\mu}\left(\bar{\varphi}^{\beta\omega}_{\mu}c
-\frac{g}{2}\varepsilon^{\rho\xi}\bar{\omega}^{\beta\omega}_{\mu}c^\rho c^\xi\right)\nonumber\cr
&&
+\tilde\Omega^\alpha\varepsilon^{\alpha\beta}\phi c^\beta
-g\varepsilon^{\alpha\beta}\tilde\tau^\alpha\phi^\beta c
-g^2\varepsilon^{\alpha\beta}\tilde\xi^\alpha\left (\varepsilon^{\beta\omega}\phi\,c^\omega c
+\frac{1}{2}\varepsilon^{\omega\rho}\phi^\beta c^\omega c^\rho\right)
-\tilde\Omega g\varepsilon^{\alpha\beta}\phi^\alpha c^\beta \nonumber\\
&&
+g\varepsilon^{\alpha\beta}\bar{\tilde{X}}^{\alpha\omega}\left(\theta^{\beta\omega}c
+\frac{g}{2}\varepsilon^{\rho\xi}\eta^{\beta\omega}c^\rho c^\xi\right)
+g\varepsilon^{\alpha\beta}\tilde{X}^{\alpha\omega}\bar{\theta}^{\beta\omega}c
+g\varepsilon^{\alpha\beta}\bar{\tilde{Y}}^{\alpha\omega}\eta^{\beta\omega}c \nonumber\\
&&%%%%%%%%%%%%%%%%%%%
\left. -g\varepsilon^{\alpha\beta}\tilde{Y}^{\alpha\omega}\left(\bar{\eta}^{\beta\omega}c
-\frac{g}{2}\varepsilon^{\rho\xi}\bar{\theta}^{\beta\omega}c^\rho c^\xi\right)
\right\}
\label{full_action}
\end{eqnarray}
%%%%%%%%%%%%%%%%%%%%%%%%%%%%%%%
The physical action \eqref{Sphys1} is reobtained from $\Sigma$ after the renormalization procedure at the limit case when  this large set of external sources and parameters achieve its physical value, namely
\begin{equation}
\{\alpha,\beta\}\to0\,,
\label{alpha_beta_zero}    
\end{equation}
\begin{equation}
\{\Omega_{\mu},\Omega^{\alpha}_{\mu},\xi^{\alpha}_{\mu},\tau^{\alpha}_{\mu},L^{\alpha},L,
\tilde\Omega,\tilde\Omega^{\alpha},\tilde\xi^{\alpha},\tilde\tau^{\alpha}\}\to0\,,
\label{BRST_sources}
\end{equation}
\begin{equation}
\{X^{\alpha}_{I},\bar{X}^{\alpha}_{I},Y^{\alpha}_{I},\bar{Y}^{\alpha}_{I},
\tilde{X}^{a}_i,\bar{\tilde{X}}^{a}_i,\tilde{Y}^{a}_i,\bar{\tilde{Y}}^{a}_i\}\to0\,,
\label{XY}
\end{equation}
\begin{equation}
\left.\bar{M}^{\alpha\beta}_{\mu\nu}\right|_{phys}=-\left.{M}^{\alpha\beta}_{\mu\nu}\right|_{phys}=\gamma^{2}\delta^{\alpha\beta}\delta_{\mu\nu}\,,\qquad
\left.\bar{N}^{\alpha\beta}_{\mu\nu}\right|_{phys}=\left.{N}^{\alpha\beta}_{\mu\nu}\right|_{phys}=0\,,
\label{MN}
\end{equation}
\begin{equation}
\left.\bar{V}^{\alpha\beta}\right|_{phys}=-\left.{V}^{\alpha\beta}\right|_{phys}=\sigma^{2}\delta^{\alpha\beta}\,,\qquad
\left.\bar{U}^{\alpha\beta}\right|_{phys}=\left.{U}^{\alpha\beta}\right|_{phys}=0
\label{UV}
\end{equation}
\begin{table}[h!]
\caption{Quantum numbers of fields and sources of the gauge sector. The nature is ``B" for bosons and ``F" for fermions.}
\begin{center}
\begin{tabular}{|l|c|c|c|c|c|c|c|c|c|c|c|c|}
\hline
\textsc{Gauge sector}$\phantom{\Bigl|}$\! &$A$&$b$&$\bar{c}$&$c$&
$\varphi$&$\bar\varphi$&$\omega$&$\bar\omega$
&$M$&$\bar{M}$&$N$&$\bar{N}$\\
\hline
\textsc{Dimension}&1&2&2&0&1&1&1&1&2&2&2&2\\
\textsc{Ghost number}&0&0&$-1$&1&0&0&1&$-1$&0&0&1&$-1$\\
$Q_8$\textsc{-Charge}&0&0&0&0&1&$-1$&1&$-1$&1&$-1$&1&$-1$\\
\textsc{Nature}&B&B&F&F&B&B&F&F&B&B&F&F\\
\hline
\end{tabular}
\end{center}
\label{Tab:gaugeSector}
\end{table}
\begin{table}[h!]
\caption{Quantum numbers of fields and sources of the matter sector. The nature is ``B" for bosons and ``F" for fermions.}
\begin{center}
\begin{tabular}{|l|c|c|c|c|c|c|c|c|c|c|c|c|c|}
\hline
\textsc{Matter sector}$\phantom{\Bigl|}$\! &$\phi$&$\theta$&$\bar{\theta}$&$\eta$&
$\bar{\eta}$&$V$&$\bar{V}$&$U$&$\bar{U}$\\
\hline
\textsc{Dimension}&1&1&1&1&1&2&2&2&2\\
\textsc{Ghost number}&0&1&$-1$&0&0&0&0&1&$-1$\\
$Q_2$\textsc{-Charge}&0&1&$-1$&1&$-1$&1&$-1$&1&$-1$\\
\textsc{Nature}&B&F&F&B&B&B&B&F&F\\
\hline
\end{tabular}
\end{center}
\label{Tab:matterSector}
\end{table}
\begin{table}[h!]
\caption{Quantum numbers of the external sources coupled to composite operators. The nature is ``B" for bosons and ``F" for fermions.}
\begin{center}
\begin{tabular}{|l|c|c|c|c|c|c|c|c|c|c|c|c|c|c|c|}
\hline
\textsc{External sources}$\phantom{\Bigl|}$\! &$\Omega$&$\tau$&$\xi$&$L$&$X$&$\bar{X}$&$Y$&$\bar{Y}$&$\tilde\Omega$&$\tilde\tau$&$\tilde\xi$&$\tilde{X}$&$\bar{\tilde{X}}$&$\tilde{Y}$&$\bar{\tilde{Y}}$\\
\hline
\textsc{Dimension}&3&3&3&4&3&3&3&3&3&3&3&3&3&3&3\\
\textsc{Ghost number}&$-1$&$-1$&$-2$&$-2$&0&$-2$&$-1$&$-1$&$-1$&$-1$&$-2$&0&$-2$&$-1$&$-1$\\
$Q_8$\textsc{-Charge}&0&0&0&0&1&$-1$&1&$-1$&0&0&0&0&0&0&0\\
$Q_2$\textsc{-Charge}&0&0&0&0&0&0&0&0&0&0&0&1&$-1$&1&$-1$\\
\textsc{Nature}&F&F&B&B&B&B&F&F&F&F&B&B&B&F&F\\
\hline
\end{tabular}
\end{center}
\label{Tab:ExtSoruces}
\end{table}

%%%%%%%%%%%%%%%%%%%%%%%%%%%%%%%%%%%%%%%%%%%%%%%%%%%%%%%%%%%%%%%%%%%%%
%%%%%%%%%%%%%%%%%%%%%%%%%%%%%%%%%%%%%%%%%%%%%%%%%%%%%%%%%%%%%%%%%%%%%
%%%%%%%%%%%%%%%%%%%%%%%%%%%%%%%%%%%%%%%%%%%%%%%%%%%%%%%%%%%%%%%%%%%%%
%%%%%%%%%%%%%%%%%%%%%%%%%%%%%%%%%%%%%%%%%%%%%%%%%%%%%%%%%%%%%%%%%%%%%
\section{Ward identities and stability for scalar matter case}

In this section we derive the large set of Ward identities fulfilled by the complete action \eqref{full_action}. These Ward identities will be the starting point for the analysis of the algebraic characterization of the most general invariant counterterm. It is easily checked that $\Sigma$ obeys the following identities: 

\begin{itemize}
\item{The Slavnov-Taylor identity:
\begin{equation}
\mathcal{S}(\Sigma)= 0
\label{stid}
\end{equation}
with
\begin{eqnarray}
\mathcal{S}(\Sigma)&\equiv&\int d^{4}x\, \left\{
\left(\frac{\delta\Sigma}{\delta\Omega^{\alpha}_{\mu}}+\frac{\delta\Sigma}{\delta\tau^{\alpha}_{\mu}}\right)\frac{\delta\Sigma}{\delta A^{\alpha}_{\mu}}
+\frac{\delta\Sigma}{\delta\Omega_{\mu}}\frac{\delta\Sigma}{\delta A_{\mu}}
+\frac{\delta\Sigma}{\delta L^{\alpha}}\frac{\delta\Sigma}{\delta c^{\alpha}}
+\frac{\delta\Sigma}{\delta L}\frac{\delta\Sigma}{\delta c}
+b^{\alpha}\frac{\delta\Sigma}{\delta\bar{c}^{\alpha}}   
+b\frac{\delta\Sigma}{\delta\bar{c}}
\right. \nonumber\\
&&+\bar\varphi^{\alpha}_I\frac{\delta\Sigma}{\delta\bar\omega^{\alpha}_I}
+\omega^{\alpha}_I\frac{\delta\Sigma}{\delta\varphi^{\alpha}_I}
-\bar{M}^{\alpha}_{\mu i}\frac{\delta\Sigma}{\delta\bar{N}^{\alpha}_{\mu I}}
+N^{\alpha}_{\mu I}\frac{\delta\Sigma}{\delta M^{\alpha}_{\mu I}}
-\left(\Omega ^\alpha_{\mu}-\tau ^\alpha_{\mu}\right)\frac{\delta\Sigma}{\delta\xi^{\alpha}_{\mu}}
+X^{\alpha}_I\frac{\delta\Sigma}{\delta Y^{\alpha}_{I}}\nonumber\\
&& +\bar{Y}^{\alpha}_{I}\frac{\delta\Sigma}{\delta\bar{X}^{\alpha}_I}
+\left(\frac{\delta\Sigma}{\delta\tilde{\Omega}^{\alpha}}+\frac{\delta\Sigma}{\delta\tilde{\tau}^{\alpha}}\right)\frac{\delta\Sigma}{\delta \phi^{\alpha}}  
+\theta^{\alpha}_i\frac{\delta\Sigma}{\delta\eta^\alpha_i}
+\bar{\eta}^{\alpha}_{i}\frac{\delta\Sigma}{\delta\bar{\theta}^{\alpha}_{i}}
-\bar{V}^\alpha_i\frac{\delta\Sigma}{\delta\bar{U}^\alpha_i}
+U^\alpha_i\frac{\delta\Sigma}{\delta{V}^\alpha_i}\nonumber\\
&&\left.-\left(\tilde{\Omega}^\alpha-\tilde{\tau}^\alpha\right)\frac{\delta\Sigma}{\delta\tilde{\xi^{\alpha}}}
   +\tilde{X}^{\alpha}_i\frac{\delta\Sigma}{\delta\tilde{Y}^{\alpha}_{i}}
   +\bar{\tilde{Y}}^{\alpha}_{i}\frac{\delta\Sigma}{\delta\bar{\tilde{X}}^{\alpha}_i}\right\}
\end{eqnarray} 
}
which is nothing but the BRST invariance of the action \eqref{full_action} when  expressed in a functional form. The BRST transformations of fields and sources can extracted from the Slavnov-Taylor identity above. For example,  the transformations of the gauge, scalar and ghost fields are given by:
\begin{eqnarray}
sA^{\alpha}_{\mu}=\frac{\delta\Sigma}{\delta\Omega^{\alpha}_{\mu}}+\frac{\delta\Sigma}{\delta\tau^{\alpha}_{\mu}}
&,&sA_{\mu}=\frac{\delta\Sigma}{\delta\Omega_{\mu}}\nonumber\\
s\phi^{\alpha}=\frac{\delta\Sigma}{\delta\tilde{\Omega}^{\alpha}}+\frac{\delta\Sigma}{\delta\tilde{\tau}^{\alpha}}
&,&s\phi=\frac{\delta\Sigma}{\delta\tilde{\Omega}}\nonumber\\
sc^{\alpha}=\frac{\delta\Sigma}{\delta L^{\alpha}}
&,&sc=\frac{\delta\Sigma}{\delta L}
\label{BRST}
\end{eqnarray}
The remaining transformations are BRST doublets. Let us also introduce, for further use, the so called  linearized Slavnov-Taylor operator  $\mathcal{B}_{\Sigma}$, defined as
\begin{eqnarray}
\mathcal{B}_{\Sigma}&=&\int d^{4}x\,\left\{ \left(\frac{\delta\Sigma}{\delta\Omega^{\alpha}_{\mu}}+\frac{\delta\Sigma}{\delta\tau^{\alpha}_{\mu}}\right)\frac{\delta}{\delta A^{\alpha}_{\mu}}
+\frac{\delta\Sigma}{\delta A^{\alpha}_{\mu}}\left(\frac{\delta}{\delta\Omega^{\alpha}_{\mu}}+\frac{\delta}{\delta\tau^{\alpha}_{\mu}}\right)
+\frac{\delta\Sigma}{\delta\Omega_{\mu}}\frac{\delta}{\delta A_{\mu}}
+\frac{\delta\Sigma}{\delta A_{\mu}}\frac{\delta}{\delta\Omega_{\mu}}
\right. \nonumber\\
&&+\frac{\delta\Sigma}{\delta L^{\alpha}}\frac{\delta}{\delta c^{\alpha}}
+\frac{\delta\Sigma}{\delta c^{\alpha}}\frac{\delta}{\delta L^{\alpha}}
+\frac{\delta\Sigma}{\delta L}\frac{\delta}{\delta c}
+\frac{\delta\Sigma}{\delta c}\frac{\delta}{\delta L}
+b^{\alpha}\frac{\delta}{\delta\bar{c}^{\alpha}}
+b\frac{\delta}{\delta\bar{c}}
+\bar\varphi^{\alpha}_I\frac{\delta}{\delta\bar\omega^{\alpha}_I}
+\omega^{\alpha}_I\frac{\delta}{\delta\varphi^{\alpha}_I}\nonumber\\
&& -\bar{M}^{\alpha}_{\mu I}\frac{\delta}{\delta\bar{N}^{\alpha}_{\mu I}}
+N^{\alpha}_{\mu I}\frac{\delta}{\delta M^{\alpha}_{\mu I}}
-\left(\Omega ^\alpha_{\mu}-\tau ^\alpha_{\mu}\right)\frac{\delta}{\delta\xi^{\alpha}_{\mu}}
+X^{\alpha}_I\frac{\delta}{\delta Y^{\alpha}_{I}}
+\bar{Y}^{\alpha}_{I}\frac{\delta}{\delta\bar{X}^{\alpha}_I}
+\left(\frac{\delta\Sigma}{\delta\tilde{\Omega}^{\alpha}}+\frac{\delta\Sigma}{\delta\tilde{\tau}^{\alpha}}\right)\frac{\delta}{\delta \phi^{\alpha}} \nonumber\\
&& +\frac{\delta\Sigma}{\delta \phi^{\alpha}}\left(\frac{\delta}{\delta\tilde{\Omega}^{\alpha}}+\frac{\delta}{\delta\tilde{\tau}^{\alpha}}\right)
+\frac{\delta\Sigma}{\delta\tilde{\Omega}}\frac{\delta}{\delta\phi}
+\frac{\delta\Sigma}{\delta\phi}\frac{\delta}{\delta\tilde{\Omega}}
+\theta^{\alpha}_i\frac{\delta}{\delta\eta^\alpha_i}
+\bar{\eta}^{\alpha}_{i}\frac{\delta}{\delta\bar{\theta}^{\alpha}_{i}}\nonumber\\
&&\left. -\bar{V}^\alpha_i\frac{\delta}{\delta\bar{U}^\alpha_i}
   +U^\alpha_i\frac{\delta}{\delta{V}^\alpha_i}
   -\left(\tilde{\Omega}^\alpha-\tilde{\tau}^\alpha\right)\frac{\delta}{\delta\tilde{\xi^{\alpha}}}
   +\tilde{X}^{\alpha}_i\frac{\delta}{\delta\tilde{Y}^\alpha_{i}}
   +\bar{\tilde{Y}}^{\alpha}_{i}\frac{\delta}{\delta\bar{\tilde{X}}^{\alpha}_i} \right\}
\end{eqnarray}

The operator $\mathcal{B}_{\Sigma}$ has the important property of being  nilpotent 
\begin{equation}
\mathcal{B}_{\Sigma} \mathcal{B}_{\Sigma} = 0 
\end{equation}

\item{The diagonal Nakanishi-Lautrup field equation:
\begin{equation}
\frac{\delta\Sigma}{\delta b}=\partial_{\mu}A_{\mu}    \label{db} 
\end{equation}
}
\item{The diagonal anti-ghost equation:
\begin{equation}
\frac{\delta \Sigma}{\delta\bar{c}}+\partial_{\mu}\frac{\delta \Sigma}{\delta\Omega_{\mu}}     = 0   \label{dantigh}
\end{equation}
}
\item{The local diagonal ghost equation:
\begin{eqnarray}
G(\Sigma)&\equiv&
\frac{\delta\Sigma}{\delta c}
+g\varepsilon^{\alpha\beta}\bigg(\bar{c}^{\alpha}\frac{\delta\Sigma}{\delta b^{\beta}}
+\bar{\omega}^\alpha_I\frac{\delta\Sigma}{\delta \bar\varphi^\beta_I}
+{\varphi}^\alpha_I\frac{\delta\Sigma}{\delta \omega^\beta_I}
+\bar{\theta}^{\alpha}_{i}\frac{\delta\Sigma}{\delta \bar\eta^{\beta}_{i}}
+{\eta}^\alpha_i\frac{\delta\Sigma}{\delta \theta^\beta_i}
+\bar{N}^{\alpha}_{\mu I}\frac{\delta\Sigma}{\delta \bar{M}^{\beta}_{\mu I}}
+{M}^\alpha_{\mu I}\frac{\delta\Sigma}{\delta {N}^{\beta}_{\mu I}}
\nonumber\\
&&
+\bar{U}^\alpha_i\frac{\delta\Sigma}{\delta \bar{V}^\beta_i}
+{V}^\alpha_i\frac{\delta\Sigma}{\delta {U}^\beta_i}
+\bar{X}^\alpha_I\frac{\delta\Sigma}{\delta \bar{Y}^\beta_I}
+{Y}^\alpha_I\frac{\delta\Sigma}{\delta {X}^\beta_I}
+\bar{\tilde{X}}^\alpha_i\frac{\delta\Sigma}{\delta \bar{\tilde{Y}}^\beta_i}
+\tilde{Y}^\alpha_i\frac{\delta\Sigma}{\delta\tilde{X}^\beta_i}
-{\xi}^{\alpha}_{\mu}\frac{\delta\Sigma}{\delta {\Omega}^{\beta}_{\mu}}
-\tilde{\xi}^{\alpha}\frac{\delta\Sigma}{\delta {\tilde{\Omega}}^{\beta}}\bigg)
\nonumber\\
&=&
-\partial_{\mu}(\partial_{\mu}\bar{c}+\Omega_{\mu})
-g\varepsilon^{\alpha\beta}\left(L^{\alpha}c^{\beta}-\tau^{a}_{\mu}A^{\beta}_{\mu}-\tilde{\tau}^{a}\phi^{\beta}
+\bar{Y}^\alpha_I\varphi^\beta_I
+{X}^\alpha_I\bar\omega^\beta_I
+\bar{X}^\alpha_I\omega^\beta_I
-{Y}^\alpha_I\bar\varphi^\beta_I\right.\nonumber\\
&&\left. +\bar{\tilde{Y}}^\alpha_i\eta^\beta_i
+\tilde{X}^\alpha_i\bar{\theta}^\beta_i
+\bar{\tilde{X}}^\alpha_i\theta^\beta_i
-\tilde{Y}^\alpha_i\bar\eta^\beta_i\right)
\label{dgh}
\end{eqnarray}
Notice that the right-hand side of eq.\eqref{dgh} is linear in the quantum fields. As such, it is a linear breaking not affected
by the quantum corrections, which is compatible with the Quantum Action Principle \cite{Piguet:1995er}. 
}

\item{The $U(1)$ residual local symmetry:
\begin{equation}
\mathcal{W}^{U(1)}\Sigma=-\partial^2b   \label{u1}
\end{equation}
where
\begin{equation}
\mathcal{W}^{U(1)}\equiv\partial_{\mu}\frac{\delta}{\delta A_{\mu}}+g\varepsilon^{\alpha\beta} \sum_{\Psi}\Psi^\alpha_{\mu}\frac{\delta}{\delta\Psi^\beta_{\mu}}
\end{equation}
being the summation over $\Psi$ a sum over all off-diagonal fields and sources, namely
\begin{eqnarray}
\Psi^{\alpha}&=&\{ A^{\alpha}_{\mu},b^{\alpha},c^{\alpha},\bar{c}^{\alpha},\phi^{\alpha},\varphi^{\alpha}_{I},\bar\varphi^{\alpha}_{I},\omega^{\alpha}_{I},\bar\omega^{\alpha}_{I}, \eta^{\alpha}_{i},\bar\eta^{\alpha}_{i},\theta^{\alpha}_{i},\bar\theta^{\alpha}_{i},
M^{\alpha}_{\mu I},\bar{M}^{\alpha}_{\mu I},N^{\alpha}_{\mu I},\bar{N}^{\alpha}_{\mu I},
V^{\alpha}_{i},\nonumber\\
&&
\bar{V}^{\alpha}_{i},U^{\alpha}_{i},\bar{U}^{\alpha}_{i},\Omega^{\alpha}_{\mu}, \tau^{\alpha}_{\mu},\xi^{\alpha}_{\mu}, L^{\alpha},\tilde\Omega^{\alpha}, \tilde\tau^{\alpha},\tilde\xi^{\alpha},X^{\alpha}_{I},\bar{X}^{\alpha}_{I},Y^{\alpha}_{I},\bar{Y}^{\alpha}_{I},
\tilde{X}^{\alpha}_{i},\bar{\tilde{X}}^{\alpha}_{i},\tilde{Y}^{\alpha}_{i},\bar{\tilde{Y}}^{\alpha}_{i}\}
\label{psi}
\end{eqnarray}
%%%
As noticed in \cite{Fazio:2001rm}, the $U(1)$ Ward identity \eqref{u1} can be obtained by anticommuting the diagonal ghost equation, eq.\eqref{dgh}, with the Slavnov-Taylor identity, eq.\eqref{stid}. This identity shows in a very clear way the fact that the diagonal component $A_\mu$ of the gauge field behaves like a $U(1)$ Abelian connection, while all off-diagonal components of the gauge and matter fields play the role of a kind of charged $U(1)$ fields, precisely like in a $QED$-type theory.
}
\item{The discrete  symmetry
\begin{equation}
\Psi^1\rightarrow\Psi^1\,,\qquad \Psi^2\rightarrow-\Psi^2 \,,\qquad
\Psi^{diag}\rightarrow-\Psi^{diag}  \label{discrete} 
\end{equation}
where $\Psi^\alpha$ and $\Psi^{diag}$ stand, respectively,  for all off-diagonal and diagonal fields and sources, namely with $\Psi^{\alpha}$ given by \eqref{psi} and
\begin{equation}
\Psi^{diag}=\{A_{\mu},b,c,\bar{c},\phi,\Omega_{\mu},L\}
\end{equation}
As pointed out in \cite{Fazio:2001rm}, this discrete symmetry plays the role of the charge conjugation with respect to the $U(1)$ Cartan subgroup of $SU(2)$. 
}
%%%
\item{The functional $\delta_{I}$'s-symmetries of gauge sector:
\begin{equation}
\mathcal{W}_I(\Sigma)\equiv \int d^{4}x\left\{\varphi^\alpha_I\frac{\delta\Sigma}{\delta\bar{c}^{\alpha}}
+c^\alpha\frac{\delta\Sigma}{\delta\bar{\varphi}^{\alpha}_I}
+\frac{\delta\Sigma}{\delta\bar{Y}^\alpha_I}\frac{\delta\Sigma}{\delta b^\alpha}
+M^\alpha_{\mu I}\frac{\delta\Sigma}{\delta\Omega^{\alpha}_{\mu}}
-Y^\alpha_{I}\frac{\delta\Sigma}{\delta L^{\alpha}}\right\}=0
\label{WIdelta1gauge}
\end{equation}
\begin{equation}
\bar{\mathcal{W}}_I(\Sigma)\equiv \int d^{4}x\left\{\bar{\omega}^\alpha_I\frac{\delta\Sigma}{\delta\bar{c}^{\alpha}}
-c^\alpha\frac{\delta\Sigma}{\delta\omega^{\alpha}_I}
+\frac{\delta\Sigma}{\delta X^\alpha_I}\frac{\delta\Sigma}{\delta b^\alpha}
-\bar{N}^\alpha_{\mu I}\frac{\delta\Sigma}{\delta\Omega^{\alpha}_{\mu}}
+\bar{X}^\alpha_{I}\frac{\delta\Sigma}{\delta L^{\alpha}}\right\}=0
\label{WIdelta2gauge}
\end{equation}
This Ward identities corresponding to $\delta_I$ and $\bar{\delta}_I$ symmetries, eqs. \eqref{delta1gauge} e \eqref{delta2gauge}, respectively, and they are responsible for the renormalization of the horizon function of the MAG \eqref{H_MAG}, as proven in \cite{Capri:2006cz}.
}
\item{The rigid $\tilde{\delta}_{i}$'s-symmetries of matter sector:
\begin{equation}
\tilde{\mathcal{W}}_i(\Sigma)\equiv \int d^{4}x\left\{\eta^\alpha_i\frac{\delta\Sigma}{\delta\bar{c}^{\alpha}}
+c^\alpha\frac{\delta\Sigma}{\delta\bar{\eta}^{\alpha}_i}
+\frac{\delta\Sigma}{\delta\bar{\tilde{Y}}^\alpha_i}\frac{\delta\Sigma}{\delta b^\alpha}
+V^\alpha_{\mu I}\frac{\delta\Sigma}{\delta\tilde{\Omega}^{\alpha}}
-\tilde{Y}^\alpha_i\frac{\delta\Sigma}{\delta L^{\alpha}}\right\}=0
\end{equation}
\begin{equation}
\bar{\tilde{\mathcal{W}}}_i(\Sigma)\equiv \int d^{4}x\left\{\bar{\theta}^\alpha_i\frac{\delta\Sigma}{\delta\bar{c}^{\alpha}}
-c^\alpha\frac{\delta\Sigma}{\delta\theta^{\alpha}_i}
+\frac{\delta\Sigma}{\delta\tilde{X}^\alpha_i}\frac{\delta\Sigma}{\delta b^\alpha}
-\bar{U}^\alpha_i\frac{\delta\Sigma}{\delta\tilde{\Omega}^{\alpha}}
+\bar{\tilde{X}}^\alpha_{I}\frac{\delta\Sigma}{\delta L^{\alpha}}\right\}=0
\end{equation}
}
This Ward identities corresponding to $\tilde{\delta}_i$ and $\bar{\tilde{\delta}}_i$ symmetries, eqs. \eqref{delta1scalar} e \eqref{delta2scalar}, respectively. This identities are completely analogous to the identities (\ref{WIdelta1gauge} -- \ref{WIdelta2gauge}). Then, this set of WI will be responsible to guarantee the renormalization of the horizon function of the matter sector \eqref{H_MAG_scalar}, or its local version \eqref{S0}.
\item{The $d_{I}$'s-symmetries of gauge sector:
\begin{eqnarray}
\mathcal{Q}_I(\Sigma)&\equiv &\int d^{4}x\left\{ \left(\omega^\alpha_I+\frac{\delta\Sigma}{\delta\bar{Y}^\alpha_I}\right)\frac{\delta\Sigma}{\delta\bar{c}^{\alpha}}
+\frac{\delta\Sigma}{\delta\bar{X}^\alpha_I}\frac{\delta\Sigma}{\delta b^\alpha}
+\frac{\delta\Sigma}{\delta L^{\alpha}}\left(\frac{\delta\Sigma}{\delta\bar{\varphi}^{\alpha}_I}-X^\alpha_{I}\right)
\right.\nonumber\\
&&\left. +c^\alpha\frac{\delta\Sigma}{\delta\bar{\omega}^{\alpha}_I}
-M^\alpha_{\mu I}\frac{\delta\Sigma}{\delta\xi^{\alpha}_{\mu}}
+N^\alpha_{\mu I}\frac{\delta\Sigma}{\delta\Omega^{\alpha}_{\mu}}\right\}=0
\end{eqnarray}
\begin{eqnarray}
\bar{\mathcal{Q}}_I(\Sigma)&\equiv &\int d^{4}x\left\{ \left(\frac{\delta\Sigma}{\delta X^\alpha_I}-\varphi^\alpha_I\right)\frac{\delta\Sigma}{\delta\bar{c}^{\alpha}}
+\frac{\delta\Sigma}{\delta Y^\alpha_I}\frac{\delta\Sigma}{\delta b^\alpha}
+\frac{\delta\Sigma}{\delta L^{\alpha}}\left(\frac{\delta\Sigma}{\delta\omega_I^\alpha}-\bar{Y}^\alpha_{I}\right)\right.\nonumber\\
&&\left. -c^\alpha\frac{\delta\Sigma}{\delta\varphi^{\alpha}_I}
-\bar{M}^\alpha_{\mu I}\frac{\delta\Sigma}{\delta\Omega^{\alpha}_{\mu}}
+\bar{N}^\alpha_{\mu I}\frac{\delta\Sigma}{\delta\xi^{\alpha}_{\mu}}\right\}=0
\end{eqnarray}
This Ward identities corresponding to $d_I$ and $\bar{d}_I$ symmetries, but also they can be obtained by anticommuting and commuting, respectively, the identities $\mathcal{W}_I$ and $\bar{\mathcal{W}}_I$ with the Slavnov-Taylor identity \eqref{stid}.
}
\item{The $\tilde{d}_{i}$'s-symmetries of matter sector:
\begin{eqnarray}
\mathcal{Q}_i(\Sigma)&\equiv &\int d^{4}x\left\{ \left(\theta^\alpha_i+\frac{\delta\Sigma}{\delta\bar{\tilde{Y}}^\alpha_i}\right)\frac{\delta\Sigma}{\delta\bar{c}^{\alpha}}
+\frac{\delta\Sigma}{\delta\bar{\tilde{X}}^\alpha_i}\frac{\delta\Sigma}{\delta b^\alpha}
+\frac{\delta\Sigma}{\delta L^{\alpha}}\left(\frac{\delta\Sigma}{\delta\bar{\eta}^{\alpha}_i}-\tilde{X}^\alpha_{i}\right)
\right.\nonumber\\
&&\left. +c^\alpha\frac{\delta\Sigma}{\delta\bar{\theta}^{\alpha}_i}
-V^\alpha_{i}\frac{\delta\Sigma}{\delta\tilde{\xi}^{\alpha}}
+U^\alpha_i\frac{\delta\Sigma}{\delta\tilde{\Omega}^{\alpha}}\right\}=0
\end{eqnarray}
\begin{eqnarray}
\bar{\mathcal{Q}}_i(\Sigma)&\equiv &\int d^{4}x\left\{ \left(\frac{\delta\Sigma}{\delta \tilde{X}^\alpha_i}-\eta^\alpha_i\right)\frac{\delta\Sigma}{\delta\bar{c}^{\alpha}}
+\frac{\delta\Sigma}{\delta\tilde{Y}^\alpha_i}\frac{\delta\Sigma}{\delta b^\alpha}
+\frac{\delta\Sigma}{\delta L^{\alpha}}\left(\frac{\delta\Sigma}{\delta\theta_i^\alpha}-\bar{\tilde{Y}}^\alpha_{i}\right)\right.\nonumber\\
&&\left. -c^\alpha\frac{\delta\Sigma}{\delta\eta^{\alpha}_i}
-\bar{V}^\alpha_i\frac{\delta\Sigma}{\delta\tilde{\Omega}^{\alpha}}
+\bar{U}^\alpha_i\frac{\delta\Sigma}{\delta\tilde{\xi}^{\alpha}}\right\}=0
\end{eqnarray}
}
This Ward identities corresponding to $\tilde{d}_I$ and $\bar{\tilde{d}}_I$ symmetries, but also they can be obtained by anticommuting and commuting, respectively, the identities $\tilde{\mathcal{W}}_i$ and $\bar{\tilde{\mathcal{W}}}_i$ with the Slavnov-Taylor identity \eqref{stid}.

\item{The rigid $\mathcal{R}$-symmetries of gauge sector:
\begin{equation}
\mathcal{R}^{(1)}_{IJ}(\Sigma)\equiv \int d^{4}x\left\{\varphi^\alpha_I\frac{\delta\Sigma}{\delta\omega^\alpha_J}
-\bar{\omega}^\alpha_J\frac{\delta\Sigma}{\delta\bar{\varphi}^\alpha_I}
+M^\alpha_{\mu I}\frac{\delta\Sigma}{\delta N^{\alpha}_{\mu J}}
+\bar{N}^\alpha_{\mu J}\frac{\delta\Sigma}{\delta\bar{M}^{\alpha}_{\mu I}}
+Y^\alpha_I\frac{\delta\Sigma}{\delta X^{\alpha}_J}
-\bar{X}^\alpha_J\frac{\delta\Sigma}{\delta\bar{Y}^{\alpha}_I}\right\}=0
\label{WIRigidGauge1}
\end{equation}
\begin{equation}
\mathcal{R}^{(2)}(\Sigma)\equiv \int d^{4}x\left\{\bar{\omega}^\alpha_I\frac{\delta\Sigma}{\delta\omega^\alpha_I}
-\bar{N}^\alpha_{\mu I}\frac{\delta\Sigma}{\delta N^{\alpha}_{\mu I}}
-\bar{X}^\alpha_I\frac{\delta\Sigma}{\delta X^{\alpha}_I}\right\}=0
\end{equation}\begin{equation}
\mathcal{R}^{(3)}(\Sigma)\equiv\int d^{4}x\left\{\bar{\omega}^\alpha_I\frac{\delta\Sigma}{\delta\varphi^\alpha_I}
-\varphi^\alpha_I\frac{\delta\Sigma}{\delta\omega^\alpha_I}
-\bar{M}^\alpha_{\mu I}\frac{\delta\Sigma}{\delta N^{\alpha}_{\mu I}}
-\bar{N}^\alpha_{\mu I}\frac{\delta\Sigma}{\delta M^{\alpha}_{\mu I}}
-\bar{X}^\alpha_I\frac{\delta\Sigma}{\delta Y^{\alpha}_I}
+\bar{Y}^\alpha_I\frac{\delta\Sigma}{\delta X^{\alpha}_I}\right\}=0
\end{equation}
}
\item{The rigid $\tilde{\mathcal{R}}$-symmetries of matter sector:
\begin{equation}
\tilde{\mathcal{R}}^{(1)}_{ij}(\Sigma)\equiv \int d^{4}x\left\{\eta^\alpha_i\frac{\delta\Sigma}{\delta\theta^\alpha_j}
-\bar{\theta}^\alpha_j\frac{\delta\Sigma}{\delta\bar{\eta}^\alpha_i}
+V^\alpha_i\frac{\delta\Sigma}{\delta U^{\alpha}_j}
+\bar{U}^\alpha_j\frac{\delta\Sigma}{\delta\bar{V}^{\alpha}_i}
+\tilde{Y}^\alpha_i\frac{\delta\Sigma}{\delta\tilde{X}^{\alpha}_j}
-\bar{\tilde{X}}^\alpha_j\frac{\delta\Sigma}{\delta\bar{\tilde{Y}}^{\alpha}_i}\right\}=0
\end{equation}
\begin{equation}
\tilde{\mathcal{R}}^{(2)}(\Sigma)\equiv \int d^{4}x\left\{\bar{\theta}^\alpha_i\frac{\delta\Sigma}{\delta\theta^\alpha_i}
-\bar{U}^\alpha_i\frac{\delta\Sigma}{\delta U^{\alpha}_i}
-\bar{\tilde{X}}^\alpha_i\frac{\delta\Sigma}{\delta\tilde{X}^{\alpha}_i}\right\}=0
\end{equation}\begin{equation}
\tilde{\mathcal{R}}^{(3)}(\Sigma)\equiv\int d^{4}x\left\{\bar{\theta}^\alpha_i\frac{\delta\Sigma}{\delta\eta^\alpha_i}
-\eta^\alpha_i\frac{\delta\Sigma}{\delta\theta^\alpha_i}
-\bar{V}^\alpha_i\frac{\delta\Sigma}{\delta U^{\alpha}_i}
-\bar{U}^\alpha_i\frac{\delta\Sigma}{\delta V^{\alpha}_i}
-\bar{\tilde{X}}^\alpha_i\frac{\delta\Sigma}{\delta\tilde{Y}^{\alpha}_i}
+\bar{\tilde{Y}}^\alpha_i\frac{\delta\Sigma}{\delta\tilde{X}^{\alpha}_i}\right\}=0
\label{WIRigidMatter3}
\end{equation}
Note that this rigid invariances in both the gauge and matter sectors, eqs. (\ref{WIRigidGauge1}--\ref{WIRigidMatter3}), as well as the discrete symmetry \eqref{discrete}, are  ``blind" with respect to the off-diagonal indices which are ``hidden" in the indices $\{I,i\}$.
}

\item{The global $U(8)$ symmetry, here written in the multi-index notation: 

\begin{eqnarray}
\mathcal{Q}_{IJ}(\Sigma)&\equiv&\int d^{4}x\,\left\{
\varphi^\alpha_I\frac{\delta}{\delta\varphi^\alpha_J}
-\bar{\varphi}^\alpha_J\frac{\delta}{\delta\bar{\varphi}^\alpha_I}
+\omega^\alpha_I\frac{\delta}{\delta\omega^\alpha_J}
-\bar{\omega}^\alpha_J\frac{\delta}{\delta\bar{\omega}^\alpha_I}
+M^\alpha_{\mu I}\frac{\delta}{\delta M^\alpha_{\mu J}}
-\bar{M}^\alpha_{\mu J}\frac{\delta}{\delta\bar{M}^\alpha_{\mu I}}
\right.\nonumber\\
&&\left.
+N^\alpha_{\mu I}\frac{\delta}{\delta N^\alpha_{\mu J}}
-\bar{N}^\alpha_{\mu J}\frac{\delta}{\delta\bar{N}^\alpha_{\mu I}}
+X^\alpha_I\frac{\delta}{\delta X^\alpha_J}
-\bar{X}^\alpha_J\frac{\delta\Sigma}{\delta\bar{X}^\alpha_I}
+Y^\alpha_I\frac{\delta}{\delta Y^\alpha_J}
-\bar{Y}^\alpha_J\frac{\delta}{\delta\bar{Y}^\alpha_I}
\right\}\Sigma =0 
\label{WI_QIJ}
\end{eqnarray}
This WI can be immediately viewed as $\mathcal{Q}_{IJ}\Sigma=0$, {\it i.e.} as a linear operator $\mathcal{Q}_{IJ}$ acting on  $\Sigma$. Then, the eigenvalues of the trace of the operator $\mathcal{Q}_{IJ}$ define the $Q_8$-charge in the Tables 1 and 3.
}
\item{The global $U(2)$ symmetry: 
\begin{eqnarray}
{\mathcal{Q}}_{ij}(\Sigma)&\equiv&\int d^{4}x\,\left\{
\eta^{\alpha}_i\frac{\delta\Sigma}{\delta\eta^{\alpha}_j}
-\bar{\eta}^{\alpha}_i\frac{\delta}{\delta\bar{\eta}^{\alpha}_j}
+\theta^{\alpha}_i\frac{\delta}{\delta\theta^{\alpha}_j}
-\bar{\theta}^{\alpha}_i\frac{\delta}{\delta\bar{\theta}^{\alpha}_j}
+U^{\alpha}_i\frac{\delta}{\delta U^{\alpha}_j}
-\bar{U}^{\alpha}_i\frac{\delta}{\delta\bar{U}^{a\alpha}_j}
\right.\nonumber\\
&&\left.
+V^{\alpha}_i\frac{\delta}{\delta V^{\alpha}_j}
-\bar{V}^{\alpha}_i\frac{\delta}{\delta\bar{V}^{\alpha}_j}
+\tilde{X}^{\alpha}_i\frac{\delta}{\delta\tilde{X}^{\alpha}_j}
-\bar{\tilde{X}}^{\alpha}_i\frac{\delta}{\delta\bar{\tilde{X}}^{\alpha}_j}
+\tilde{Y}^{\alpha}_i\frac{\delta}{\delta\tilde{Y}^{\alpha}_j}
-\bar{\tilde{Y}}^{\alpha}_i\frac{\delta}{\delta\bar{\tilde{Y}}^{\alpha}_j}
\right\} \Sigma =0
\label{WI_Qij}
\end{eqnarray}
}
Analogously to the previous WI, the equation above can be written as $\mathcal{Q}_{ij}\Sigma=0$ and the trace of the linear operator $\mathcal{Q}_{ij}$ defines the $Q_2$-charge in the Tables 2 and 3.
\end{itemize}

\subsection{Renormalization factors}

The corresponding counterterm, \textit{i.e.}, the most general integrated local polynomial in the fields and sources, with dimension four and ghost number zero, compatible with all symmetries of the action, that can be freely added at order $\epsilon$ in the perturbative expansion, is given by
\begin{equation}
\Sigma_{c.t.}=\Sigma_{0}+   \mathcal{B}_{\Sigma} \Delta^{-1}   
 \label{CT_gamma_zero}
\end{equation}
where $\Sigma_{0}$ stands for the nontrivial part of the cohomolgy of the operator $\mathcal{B}_{\Sigma}$, being given by 
\begin{equation}
\Sigma_{0}=a_0 S_{YM}+\int d^4x\left(a_1\frac{m_{\phi}^2}{2}\phi^a\phi^a
+a_{2} \frac{\lambda}{4!}(\phi^a\phi^a)^2 \right) 
\end{equation}
and $\Delta^{(-1)}$ is given by an integrated local polynomial in the fields with dimension $4$, ghost number $(-1)$ and with vanishing $\mathcal{Q}_8$ and $\mathcal{Q}_2$ charges. Taking into account the full set of symmetries of the last section, one can write $\Delta^{(-1)}$ as
\begin{eqnarray}
\Delta^{-1}&=&\int d^4x\left\{(a_3+a_4)(\Omega^\alpha_{\mu}A^\alpha_{\mu}+g\varepsilon^{\alpha\beta}\xi^\alpha_{\mu}A^\beta_{\mu}c)
+(a_4+a_5)\xi^\alpha_{\mu}(D^{\alpha\beta}_{\mu}c^\beta)-a_6\bar{c}^\alpha D^{\alpha\beta}_{\mu}A^\beta_{\mu}
+a_7c^\alpha L^\alpha
\right.\nonumber\\
&&
+(a_3-a_5+a_7)(\bar{N}^\alpha_{\mu I}D^{\alpha\beta}_{\mu}\varphi^\beta_I-M^\alpha_{\mu I}D^{\alpha\beta}_{\mu}\bar{\omega}^\beta_I)
-(a_6+a_7)\bar{\omega}^\alpha_I\mathcal{M}^{\alpha\beta}\varphi^\beta_I+a_9\chi\bar{N}^\alpha_{\mu I}M^\alpha_{\mu I}
\nonumber\\
&& +(a_{10}+a_{11})(\tilde{\Omega}^\alpha\phi^a+g\varepsilon^{\alpha\beta}\tilde{\xi}^\alpha\phi^\beta c-\tilde{\Omega}\phi)
+(-a_{11}+a_12)g\varepsilon^{\alpha\beta}\tilde{\xi}^\alpha\phi c^\beta
\nonumber\\
&&
+(a_{10}-a_{12}+a_7)g\varepsilon^{\alpha\beta}(-\bar{U}^\alpha_i\phi\eta^\beta_i +V^\alpha_i\phi\bar{\theta}^\beta_i)
-a_7\bar{\theta}^\alpha_i\mathcal{M}^{\alpha\beta}\eta^\beta_i+a_{13}\tilde{\chi}\bar{U}^\alpha_iV^\alpha_i
\nonumber\\
&&
-\alpha\left[ a_{14}(\bar{c}^\alpha b^\alpha-g\varepsilon^{\alpha\beta}\bar{c}^\alpha\bar{c}^\beta c)
-(a_7+2a_{14})g^2\bar{c}^\alpha c^\alpha(\varphi^\beta_I\bar{\omega}^\beta_I +\eta^\beta_i\bar{\theta}^\beta_i)\right]\nonumber\\
&& 
+\alpha(a_7+a_{14})g^2\left[(2\bar{\omega}^\alpha_I\varphi^\alpha_I+\eta^\alpha_i\bar{\theta}^\alpha_i)
(\bar{\varphi}^\beta_J\varphi^\beta_J-\bar{\omega}^\beta_J\omega^\beta_J)
+(2\bar{\theta}^\alpha_i\eta^\alpha_i+\varphi^\alpha_I\bar{\omega}^\alpha_I)
(\bar{\eta}^\beta_j\eta^\beta_j-\bar{\theta}^\beta_j\theta^\beta_j)\right]\nonumber\\
&&
\left. +\beta(a_7+a_{15})\left[\varepsilon^{\alpha\beta}\phi\phi^\alpha\bar{c}^\beta
+g\phi^\alpha\phi^\alpha(\varphi^\beta_I\bar{\omega}^\beta_I+\eta^\beta_i\bar{\theta}^\beta_i)
-g\phi^\alpha\phi^\beta(\varphi^\beta_I\bar{\omega}^\alpha_I+\eta^\beta_i\bar{\theta}^\alpha_i)
+g\phi^2(\varphi^\beta_I\bar{\omega}^\beta_I-\eta^\beta_i\bar{\theta}^\beta_i)
\right]
\right\}
\end{eqnarray}
where $\{a_k\}_{k=1}^{15}$ are independent arbitrary coefficients. The counterterm \eqref{CT_gamma_zero} can be reabsorbed in the classical action by a multiplicative renormalization  of the fields, sources and parameters:
\begin{equation}
\Sigma[\mathcal{F}_0,\mathcal{J}_0]+O(\epsilon^{2})=\Sigma[\mathcal{F},\mathcal{J}]+\epsilon\,\Sigma_{\mathrm{CT}}
\end{equation}
where, the label ``0'' indicates a bare (nonrenormalized) quantity, $\epsilon$ is the expansion parameter,  $\mathcal{F}$ stands for the fields and $\mathcal{J}$ stands for  the external sources and parameters. By convention we choose the renormalization factors as
\begin{eqnarray}
\mathcal{F}_0&=&Z^{1/2}_{\mathcal{F}}\,\mathcal{F}=\left(1+\frac{\epsilon}{2}\,z_{\mathcal{F}}\right)\mathcal{F}\nonumber\\
\mathcal{J}_0&=&Z_{\mathcal{J}}\,\mathcal{J}=\left(1+\epsilon\,z_{\mathcal{J}}\right)\mathcal{J}
\end{eqnarray} 
where the coefficients $\{z_{\mathcal{F}},z_{\mathcal{J}}\}$ are certain linear combinations of $\{a_k\}$. By direct inspection, one can find that
\begin{eqnarray}
(Z_A^{off})^{1/2}&=&1+\epsilon\left(\frac{a_0}{2}+a_3+a_4\right)\\
Z_g&=&1-\epsilon\frac{a_0}{2}\\
(Z_{\phi}^{off})^{1/2}&=&1+\epsilon(a_{10}+a_{11})\\
(Z_{\phi}^{\text{diag}})^{1/2}&=&1-\epsilon(a_{11}+a_{12})\\
Z_{m_{\phi}}&=&1+\epsilon\frac{a_1}{2}\\
Z_{\lambda}&=&1+\epsilon a_2\\
(Z_b^{off})^{1/2}&=&1-\epsilon\left(\frac{a_0}{2}+a_6\right)\\
Z_{\alpha}&=&1+\epsilon(a_0+2a_6-2a_{14})\\
(Z_{\bar{c}}^{off})^{1/2}=(Z_c^{off})^{1/2}&=&1-\epsilon\left(\frac{a_6+a_7}{2}\right)\\
(Z_c^{diag})^{1/2}&=&1+\epsilon\left(\frac{a_7-a_6}{2}\right)\\
Z_{\bar{\omega}}^{1/2}&=&1-\epsilon\left(\frac{a_0}{2}+a_7\right)\\
Z_{\omega}^{1/2}&=&1+\epsilon\left(\frac{a_0}{2}-a_7\right)\\
Z_{\bar{M}}=Z_M&=&1-\epsilon\left(\frac{a_6-a_7}{2}-a_3-a_4\right)\\
Z_N&=&1+\epsilon\left(\frac{a_0}{2}-a_3+a_5\right)\\
Z_{\bar{N}}&=&1+\epsilon\left(\frac{a_0}{2}-a_3+a_5+a_6-a_7\right)\\
Z_{\chi}&=&1+\epsilon\left(2a_3-a_6+a_7-a_{9}\right)\\
Z^{1/2}_{\bar{\eta}}=Z^{1/2}_{\eta}&=&1-\epsilon\frac{a_7}{2}\\
Z^{1/2}_{\bar{\theta}}&=&1+\epsilon\left(\frac{a_0-a_6}{2}-a_7\right)\\
Z^{1/2}_{\theta}&=&1-\epsilon\left(\frac{a_0-a_6}{2}\right)\\
Z_{\bar{V}}=Z_V&=&1+\epsilon\left(\frac{a_0+a_7}{2}+a_{10}-a_{12}\right)\\
Z_{\bar{U}}&=&1+\epsilon\left(a_0-a_{10}+a_{12}-\frac{a_6}{2}\right)\\
Z_{U}&=&1+\epsilon\left(a_{10}-\tilde{a}_4+2a_7+\frac{a_6}{2}\right)\\
Z_{\tilde{\chi}}&=&1-\epsilon(a_0+2a_{10}-2a_{12}+a_7-a_{13})
\end{eqnarray}
and
\begin{eqnarray}
(Z_A^{diag})^{1/2}&=&(Z_b^{diag})^{-1/2}=Z_g^{-1}\\
(Z_{\bar{c}}^{diag})^{1/2}&=&(Z_{c}^{diag})^{-1/2}\\
Z_{\bar{\varphi}}^{1/2} &=&Z_{\varphi}^{1/2}=(Z_{\bar{c}}^{off})^{1/2}
\end{eqnarray}
%%%%%%%%%%%%%%%%%%%%%%%%%%%%%%%%%%%%%%%%%%%%%%%%
Finally, we note that the non-renormalization theorem of the maximal Abelian gauge \cite{Fazio:2001rm}, namely  $Z_g(Z_A^{diag})^{1/2}=1$ remains true in the presence of the horizon matter function, extending the result finding in \cite{Capri:2015pxa}. This ends the proof of the multiplicative renormalization of the $SU(2)$ Gribov-Zwanziger action in the MAG with confining scalar matter \eqref{Sphys1} which is the physical limit, \textit{see} (\ref{alpha_beta_zero}--\ref{UV}), of the extended action \eqref{full_action}. Notice that in the physical limit we can recover the renormalization factors of Gribov's parameters; then we have that
\begin{eqnarray}
Z_{\gamma}&=&Z_M|_{\text{phys}} = (Z_{c}^{\text{diag}})^{1/2}(Z_{A}^{\text{off}})^{1/2}Z_g \\
Z_{\sigma}&=&Z_V|_{\text{phys}} = Z_g Z_{\bar{\eta}}^{-1/2}(Z_{\phi}^{\text{off}})^{1/2}(Z_{c}^{\text{diag}})^{1/2}
\end{eqnarray}
%%%%%%%%%%%%%%%%%%%%%%%%%%%%%%%%%%%%%%%%%%%%%%%%
%%%%%%%%%%%%%%%%%%%%%%%%%%%%%%%%%%%%%%%%%%%%%%%%

\section{Conclusion}

In this work we have addressed the issue of the all orders perturbative  renormalization of the $SU(2)$ Gribov-Zwanziger model in the maximal Abelian gauge in the presence of confined scalar matter fields in the adjoint representation as well as fermion matter in the fundamental representation. Following the conjecture of  universal coupling for Faddeev-Popov operator to any coloured field, as proposed in \cite{Capri:2014bsa}, an additional Gribov-like term in the matter  sector is implemented in order to compel the confinement character of the matter field, which shares great similarity with the horizon function introduced in the pure gauge sector,  providing  that a similar picture can be consistently achieved in the matter case via the Gribov-Zwanziger framework. Due to the non-linearity of the gauge fixing condition  a new quartic interaction terms in  scalar matter fields, off-diagonal Faddeev-Popov ghosts and Zwanziger-like localizing fields are required for renormalizabilty. These new terms are BRST-invariant, as expressed by eq.\eqref{Sigma_quartic}, and proportional to a new gauge-like parameter $\beta$, generalizing the main result reported in \cite{Capri:2015pxa} to the Gribov context.  Moreover, the most remarkable fact is the existence of a new set of symmetries relating the  auxiliary localizing Zwanziger-like fields in the matter sector and the Faddeev-Popov fields, eqs. (\ref{delta1scalar}--\ref{delta4scalar}), in perfect analogy to the symmetries in the pure gauge sector reported in \cite{Capri:2006cz,Capri:2008ak},  which allow us to keep under control the ultraviolet finiteness  of the new horizon-like term in the matter through the existence of a set of associated Ward identities..

The analysis of the all orders perturbative renormalizability of the maximal Abelian gauge in presence of matter fields is the first necessary step towards the investigation of the non-perturbative effects of the Gribov copies, which deeply affect the maximal Abelian gauge \cite{Capri:2006cz,Capri:2008ak,Capri:2008vk,Capri:2010an}. We underline that the requirements of localizability and renormalizability are unavoidable in order to have at our disposal a consistent computational framework. Besides, although the proof of the renormalizability given here refers to the gauge group $SU(2)$, it can be easily generalized to other gauge groups as well as to other representations of the scalar fields. 

 The resulting local form of the  full action is obtained by the introduction of auxiliary Zwanziger-like fields which, as in the case of the localizing Zwanziger fields of the pure gauge sector, develop their own dynamics giving rise to the formation of diemnsion two  condensates, as explicitly checked through one-loop computations in \cite{Capri:2017abz}. Moreover, the condensates arising in the matter sector can be taken into account through an effective action which looks much alike the refined Gribov-Zwanziger action which accounts for the existence of similar condensates in the gluon sector. The inclusion of the dimension two operators is straightforward and don't spoil the renormalization of the model \cite{Capri:2008ak}.

Finally, the inclusion of the usual Dirac action for spinors does not pose any additional problem. In the same way as before, the renormalizability is guaranteed by a new set of Ward identities analogous to the above-mentioned, as showed in the eqs. (\ref{delta1fermionic}--\ref{delta4fermionic}) at the Appendix \ref{AppFerCase}. Also, unlike the case of scalar matter fields, BRST invariance and power counting do not allow for additional interaction terms between spinors and Faddeev-Popov ghosts.

%%%%%%%%%%%%%%%%%%%%%%%%%%%
\section*{Acknowledgments}
%%%%%%%%%%%%%%%%%%%%%%%%%%%
The Coordena{\c{c}}{\~{a}}o de Aperfei{\c{c}}oamento de Pessoal de
N{\'{\i}}vel Superior (CAPES)  is gratefully acknowledged. 

\appendix
 
\section{Renormalization for fermion matter case}
\label{AppFerCase}

The inclusion of the usual Dirac action for spinors does not pose any additional problem. In fact, as in the Landau gauge \cite{Capri:2014bsa,Capri:2014fsa}, the extension to the case of  fermion matter in the MAG is immediate and its renormalizability follows by analogy with the renormalization of the scalar matter case. The same arguments presented for the scalar case can be repeat in the spinorial case, providing that an analogous set of Ward identities can be established.

In complete analogy to the Subsection \ref{SubSecConfScalarMatter}, considering the case which spinor matter is present, then we added the Dirac action in the  fundamental  representation (whose indixes are represented by lowercase Latin letters) minimally coupled to gauge field, \textit{i.e.}
\begin{equation}
S_{\text{spinor}}=\int d^4x \left[\bar{\psi}^i_{\hat{\alpha}}(\gamma_{\mu})_{\hat{\alpha}\hat{\beta}}D_{\mu}^{ij}
 \psi^{j\hat{\beta}}-m_{\psi}\bar{\psi}^i_{\hat{\alpha}}\psi^{i\hat{\alpha}}\right]
\label{Sspinor}
\end{equation}
where covariant derivative is in fundamental representation. Note that at this case the circumflexed Greek indixes are spinorial, they do not correspond to the off-diagonal components in a Cartan decomposition, as the case of no-circumflexed Greek letters.  The confining character of spinor matter is implemented by adding an horizon function in this sector which coupled the inverse of Faddeev-Popov operator \eqref{offop} to off-diagonal generators in the fundamental representation, according to
\begin{equation}
H_{\text{spinor}}(\psi)=-g^2\int d^4xd^4y\,\bar{\psi}^i_{\hat{\alpha}}(x)(T^\alpha)^{ij}(\mathcal{M}^{-1})^{\alpha\beta}(x,y)(T^\beta)^{jk}\psi^{k\hat{\alpha}}(y)
\label{Hnlocal}
\end{equation}
The complete non-local IR matter action in this case is given by
\begin{equation}
S_{\text{matter}}=S_{\text{spinor}}+M^3H_{\text{spinor}}
\label{Sphys2}
\end{equation}
The parameter $M$ is analogous to the Gribov parameter for the case of spinorial matter. Evidently, in same way as before, the non-local horizon function \eqref{Hnlocal} can be cast in local form by the usual method, introducing the  fields $(\bar{\eta}^{\alpha i}_{\hat{\alpha}},\eta^{\alpha i\hat{\alpha}})$ and $(\bar{\lambda}^{\alpha i}_{\hat{\alpha}},\lambda^{\alpha i\hat{\alpha}})$, anticommuting and commuting, respectively, where we change the notation of the Zwanziger fields in order to keep the harmony with \cite{Capri:2014bsa,Capri:2014fsa,Capri:2017abz}. Thus, the local action of quark matter fields coupled with the gauge sector in a non-perturbative way is expressed as
\begin{equation}
H^{\text{local}}_{\text{spinor}}= \int d^4x \left\{ 
\bar{\lambda}^{\alpha i}_{\hat{\alpha}}\mathcal{M}^{\alpha\beta}\lambda^{\beta i \hat{\alpha}}
-\bar{\eta}^{\alpha i}_{\hat{\alpha}}\mathcal{M}^{\alpha\beta}\eta^{\beta i \hat{\alpha}}
-gM^{3/2}\left[\bar{\lambda}^{\alpha i}_{\hat{\alpha}} (T^\alpha)^{ij} \psi^{j\hat{\alpha}}
-\bar{\psi}^{i}_{\hat{\alpha}} (T^\alpha)^{ij} \lambda^{\alpha j\hat{\alpha}}\right]
\right\}
\label{HlocalSpinor}
\end{equation}
After localization, we can see that full matter action \eqref{Sphys2} exhibits a soft breaking of the nilpotent BRST symmetry 
\begin{eqnarray}
&s\psi^{i}_{\hat{\alpha}}=-ig(T^{a})^{ij}\,c^{a}\psi^{j}_{\hat{\alpha}}\,,\qquad
s\bar\psi^{i}_{\hat{\alpha}}=-ig\,\bar\psi^{j}_{\hat{\alpha}}(T^{a})^{ji}c^{a}&\nonumber\\
&s\bar\eta^{\alpha i}_{\hat{\alpha}}=\bar\lambda^{\alpha i}_{\hat{\alpha}}\,,\qquad s\bar\lambda^{\alpha i}_{\hat{\alpha}}=0\,,\qquad
s\lambda^{ai}_{\hat{\alpha}}=\eta^{ai}_{\hat{\alpha}}\,,\qquad s\eta^{ai}_{\hat{\alpha}}=0& 
\label{brstf} 
\end{eqnarray} 
due to the presence of Gribov-like parameter $M$. Note the difference between the Latin and Greek indices in the above transformations. As usual, we write the BRST-exact form for this case by  introducing a quartet of sources $\left( \bar{U}^{ij}_{\hat{\alpha}\hat{\beta}},U^{ij}_{\hat{\alpha}\hat{\beta}},\bar{V}^{ij}_{\hat{\alpha}\hat{\beta}},V^{ij}_{\hat{\alpha}\hat{\beta}}\right)$, namely\footnote{\textit{Cf}. the second line in \eqref{Sources_NMUV} and the eq. \eqref{ActSUV}.}
\begin{eqnarray}
S_{UV}&=&s\int d^{4}x\,\left\{
\bar{U}^{jk}_{\hat{\alpha}\hat{\beta}}\,\bar\psi^{i\hat{\alpha}}g(T^{\alpha})^{ij}\lambda^{\alpha k\hat{\beta}}
+V^{jk}_{\hat{\alpha}\hat{\beta}}\,\bar\eta^{\alpha k\hat{\beta}}g(T^{\alpha})^{ij}\psi^{j\hat{\alpha}}
+\zeta m_{\psi}\,\bar{U}^{ij}_{\hat{\alpha}\hat{\beta}}V^{ij}_{\hat{\alpha}\hat{\beta}}
\right\}\nonumber\\
%%%%%%%%%%%%%%%%%%%%%%%%%%%%%%%%%%%%%%%%%%%%%%%%%%%%%
&=&\int d^4x\left\{\bar{V}^{jk}_{\hat{\alpha}\hat{\beta}}\,\bar\psi^{i\hat{\alpha}}g(T^{\alpha})^{ij}\lambda^{\alpha k\hat{\beta}}
+\bar{U}^{jk}_{\hat{\alpha}\hat{\beta}}\left[ig^{2}(T^{\alpha})^{ij}(T^{b})^{\ell i}\bar\psi^{\ell\hat{\alpha}}c^{b}\lambda^{\alpha k\hat{\beta}}
+\bar\psi^{i\hat{\alpha}}g(T^{\alpha})^{ij}\eta^{\alpha k\hat{\beta}}\right]
\right.\nonumber\\
&&\quad +V^{ik}_{\hat{\alpha}\hat{\beta}}\left[\bar\lambda^{\alpha k\beta}g(T^{\alpha})^{ij}\psi^{j\hat{\alpha}}-ig^{2}\bar\eta^{\alpha k\hat{\beta}}(T^{\alpha})^{ij}(T^{b})^{j\ell}c^{b}\psi^{\ell\hat{\alpha}}\right]
+U^{ik}_{\hat{\alpha}\hat{\beta}}\bar\eta^{\alpha k\hat{\beta}}g(T^{\alpha})^{ij}\psi^{j\hat{\alpha}}\nonumber\\
&&\quad +\zeta m_{\psi}\,\left(\bar{V}^{ij}_{\hat{\alpha}\hat{\beta}}V^{ij}_{\hat{\alpha}\hat{\beta}}
-\bar{U}^{ij}_{\alpha\beta}U^{ij}_{\hat{\alpha}\hat{\beta}}\right)\biggr\}
\label{SUV}
\end{eqnarray} 

The last term in expression above, proportional to  the dimensionless coefficient $\zeta$,  is a vacuum term allowed  by power-counting. The term proportional to Gribov parameter $M$ in \eqref{HlocalSpinor} is recovered from  the invariant action  $S_{UV}$ when the external sources attain the so-called physical value, {\it i.e.} 

\begin{equation}
V^{ij}_{\alpha\beta}\Bigl|_{\mathrm{phys}}=\bar{V}^{ij}_{\alpha\beta}\Bigl|_{\mathrm{phys}}=M^{3/2}\delta^{ij}\delta_{\alpha\beta}\,,\qquad
U^{ij}_{\alpha\beta}\Bigl|_{\mathrm{phys}}=\bar{U}^{ij}_{\alpha\beta}\Bigl|_{\mathrm{phys}}=0
\end{equation}

A composite index $\hat{I} \equiv \{i, \hat{\alpha}\}$ (a combination of fundamental representation and spinorial indixes) can be introduced, which relies on an exact $U(8)$ symmetry. Therefore a new symmetry arise  in perfect analogy with (\ref{delta1scalar}-\ref{delta4scalar}), which relate Zwanziger-like spinorial sector with the Faddeev-Popov sector fields, namely
\begin{itemize}
\item{$\delta_{\hat{I}}$-symmetry
\begin{equation}
\delta_{\hat{I}}\bar{c}^\alpha=\lambda^\alpha_{\hat{I}}\,,\quad\delta_{\hat{I}}\bar{\lambda}^\alpha_j=\delta_{\hat{I}\hat{J}}c^\alpha\,,\quad
\delta_{\hat{I}}b^\alpha =g\varepsilon^{\alpha\beta}\lambda^\beta_{\hat{I}}c\,,\quad
\delta_{\hat{I}}\bar{J}^i_{\hat{\alpha}} =V^i_{\hat{\alpha}\hat{I}}
\label{delta1fermionic}
\end{equation}
}
\item{$\bar{\delta}_{\hat{I}}$-symmetry
\begin{equation}
\bar{\delta}_{\hat{I}}\bar{c}^\alpha=\bar{\eta}^\alpha_{\hat{I}}\,,\quad
\bar{\delta}_{\hat{I}}\eta^\alpha_{\hat{J}}=-\delta_{\hat{I}\hat{J}} c^\alpha\,,\quad
\bar{\delta}_{\hat{I}}b^\alpha =g\varepsilon^{\alpha\beta}\bar{\eta}^\beta_{\hat{I}}c\,,\quad
\bar{\delta}_{\hat{I}}J^i_{\hat{\alpha}} =-\bar{U}^{i}_{\hat{\alpha}\hat{I}}
\label{delta2fermionic}
\end{equation}
}
\item{$d_{\hat{I}}$-symmetry
\begin{eqnarray}
d_{\hat{I}}\bar{c}^\alpha =\eta^\alpha_{\hat{I}}+g\varepsilon^{\alpha\beta}\lambda^\beta_{\hat{I}}c\,,\quad
d_{\hat{I}}\bar{\lambda}^\alpha_{\hat{J}}=\delta_{\hat{I}\hat{J}} g\varepsilon^{\alpha\beta}c^\beta c\,,\quad
d_{\hat{I}}b^\alpha =g\varepsilon^{\alpha\beta}\eta^\beta_{\hat{I}}c
+\frac{g^2}{2}\varepsilon^{\alpha\beta}\varepsilon^{\gamma\delta}\lambda_{\hat{I}}^\beta c^\gamma c^\delta\,,
\nonumber\\
d_{\hat{I}}\bar{\eta}^\alpha_{\hat{J}}=\delta_{\hat{I}\hat{J}}c^\alpha \,,\quad
d_{\hat{I}}J^i_{\hat{\alpha}} =U^{i}_{\hat{\alpha}\hat{I}}\,,\quad
d_{\hat{I}}\xi^{i}_{\hat{\alpha}} =-V^{i}_{\hat{\alpha}\hat{I}}\qquad\qquad
\label{delta3fermionic}
\end{eqnarray}
}
\item{$\bar{d}_{\hat{I}}$-symmetry
\begin{eqnarray}
\bar{d}_{\hat{I}}\bar{c}^\alpha =-\eta^\alpha_{\hat{I}}+g\varepsilon^{\alpha\beta}\bar{\theta}^\beta_{\hat{I}}c\,,\quad
\bar{d}_{\hat{I}}\theta^\alpha_{\hat{J}}=\delta_{\hat{I}\hat{J}}g\varepsilon^{\alpha\beta}c^\beta c\,,\quad
\bar{d}_{\hat{I}}b^\alpha =-g\varepsilon^{\alpha\beta}\bar{\eta}^\beta_{\hat{I}}c
+\frac{g^2}{2}\varepsilon^{\alpha\beta}\varepsilon^{\gamma\delta}\bar{\theta}_{\hat{I}}^\beta c^\gamma c^\delta\,, \nonumber\\
\bar{d}_{\hat{I}}\eta^\alpha_j=-\delta_{\hat{I}\hat{J}}c^\alpha\,,\quad
\bar{d}_{\hat{I}}\bar{J}^{i}_{\hat{\alpha}}=-\bar{V}^{i}_{\hat{\alpha}\hat{I}}\,,\quad
\bar{d}_{\hat{I}}\bar{\xi}^{i}_{\hat{\alpha}}=\bar{U}^{i}_{\hat{\alpha}\hat{I}}\qquad\qquad
\label{delta4fermionic}
\end{eqnarray}
}
\end{itemize}

where, because the non-linearity of \eqref{brstf} for spinors,  we introduce external s-invariant sources $(\bar{J}^{i\hat{\alpha}},J^{i}_{\hat{\alpha}})\equiv (\bar{J}_{\hat{I}},J_{\hat{I}})$ and $(\bar{K}^{i\hat{\alpha}},K^{i}_{\hat{\alpha}})\equiv (\bar{K}_{\hat{I}},K_{\hat{I}})$ coupled to the nonlinear BRST transformations in the off-diagonal and diagonal Faddeev-Popov ghosts, in such a way that $s\xi^{i}_{\hat{\alpha}}=-(J^{i}_{\hat{\alpha}}-K^{i}_{\hat{\alpha}})$ and similarly for a source $\bar{\xi}^i_{\hat{\alpha}}$\footnote{Cf. the equations \eqref{ExtSoruces01} and \eqref{ExtSoruces02}.}. Also, unlike the case of scalar matter fields, BRST invariance and power counting do not allow for additional interaction terms between spinors and Faddeev-Popov ghosts, like the $\beta$-terms in \eqref{Sigma_quartic} or something of the kind, which is the biggest difference in relation to scalar matter case.

\bibliographystyle{plain}
\bibliography{biblio}

\end{document}